\documentclass[a4paper,11pt]{article}
\usepackage{jcappub} 
\usepackage[T1]{fontenc} 
\usepackage{amsmath,amssymb,amsfonts,amsthm}
\usepackage{physics}
\newcommand{\GCD}{\hat{\nabla}{}}  % General Connection Covariant Derivative

\newcommand{\LCg}{\mathring{\Gamma}{}} % Levi-Civita connection of the arbitrary metric g
\newcommand{\CDg}{\mathring{\nabla}{}}   % Levi-Civita covariant derivative of the arbitrary metric $g$
\newcommand{\Rg}{\mathring{R}{}}   % Levi-Civita curvature of the arbitrary metric $g$

\newcommand{\dimM}{\mathtt{n}}   % Dimension of the manifold
\newcommand{\ii}{\mathrm{i}}     % Imaginary unit
\renewcommand{\order}[1]{O(#1)}
\newcommand{\CPsi}{\overline{\Psi}}

\title{Bootstrapping gravity and its extension to metric-affine theories}

\author[a,b]{Adri\`a Delhom,}
\author[c]{Gerardo Garc\'ia-Moreno,}
\author[b]{Manuel Hohmann,}
\author[b]{Alejandro Jim\'enez Cano}
\author[b,d]{and Tomi S. Koivisto}

\affiliation[a]{Department of Physics and Astronomy, Louisiana State University, Baton Rouge, Louisiana 70803, USA}
\affiliation[b]{Laboratory of Theoretical Physics, Institute of Physics, University of Tartu, W. Ostwaldi 1, 50411 Tartu, Estonia}
\affiliation[c]{Instituto de Astrof\'{\i}sica de Andaluc\'{\i}a (IAA-CSIC), Glorieta de la Astronom\'{\i}a, 18008 Granada, Spain}
\affiliation[d]{National Institute of Chemical Physics and Biophysics, R\"avala pst. 10, 10143 Tallinn, Estonia}

\emailAdd{adria.delhom@gmail.com}
\emailAdd{ggarcia@iaa.es}
\emailAdd{manuel.hohmann@ut.ee}
\emailAdd{alejandro.jimenez.cano@ut.ee}
\emailAdd{tomi.koivisto@ut.ee}

\abstract{In this work we study diffeomorphism-invariant metric-affine theories of gravity from the point of view of self-interacting field theories on top of Minkowski spacetime (or other background). We revise how standard metric theories couple to their own energy-momentum tensor, and discuss the generalization of these ideas when torsion and nonmetricity are also present. We review the computation of the corresponding currents through the Hilbert and canonical (Noether) prescriptions, emphasizing the potential ambiguities arising from both. We also provide the extension of this consistent self-coupling procedure to the vielbein formalism, so that fermions can be included in the matter sector. In addition, we clarify some subtle issues regarding previous discussions on the self-coupling problem for metric theories, both General Relativity and its higher derivative generalizations. We also suggest a connection between Lovelock theorem and the ambiguities in the bootstrapping procedure arising from those in the definition of conserved currents.}

\begin{document}
\maketitle
\flushbottom

%------------------------------------------------
\section{Introduction}
\label{Sec:Introduction}
%------------------------------------------------

Our modern understanding of gravitation is through Einstein's theory of General Relativity (GR), which naturally provides a geometric description of gravitational phenomena. It is often emphasized that one of the key properties of General Relativity is that it is invariant under diffeomorphisms, namely, that it does not contain any kind of privileged background structures. The arena where everything occurs, spacetime, becomes itself dynamical. The rest of the fundamental interactions, unified in the Standard Model of Particle Physics are described as Quantum Field Theories on top of a fixed Minkowskian background. This seems to pose a huge bridge when thinking about merging both formalisms together. However, there is a way of reformulating GR as a field theory on Minkowski spacetime. This was first worked out by Rosen~\cite{Rosen1940a,Rosen1940b}, who showed that GR can be rewritten as a field theory on top of Minkowski spacetime. This field carries the degrees of freedom corresponding to the self-interacting massless spin-2 representation of the Poincaré Group, {\it i.e.} the graviton~\cite{Wigner1939,Fierz1939,Weinberg1995}.

Later on, Gupta~\cite{Gupta1954} suggested that the structure of GR is such that it would be the only possible non-linear extension of the linearized theory, suggesting that Rosen's construction of the interacting theory of a massless spin-2 field was unique. There were many works addressing this problem and the uniqueness of the construction (see {\it e.g.} \cite{Kraichnan1955} and \cite{Huggins1962}). Among them, we wish to highlight Feynman's contributions \cite{Feynman1996}, where he showed that the unique non-linear self-consistent theory that one could build starting from the linear massless spin-2 field theory is one in which, order by order, the field couples to its own energy-momentum tensor. This approach is today called a \textit{bootstrapping} procedure. The approach taken in this line of work is that, after adding up the infinite series of self-consistent interactions, one would end up with a non-linear theory which ideally should be uniquely GR. Later, Deser showed that GR is a solution to consistency problem by implementing a similar idea in the first order formalism \cite{Deser1970} (see also~\cite{Ortin2015} for a clear presentation of this approach), although uniqueness still slipped away.

In a different line, Wald proposed another approach based on a different guiding principle, which led to similar conclusions \cite{Wald1986,Wald1986b}. To be more specific, instead of focusing on the self-coupling to its own energy-momentum tensor, Wald insisted only on preserving the linear Bianchi identities associated with the gauge invariance of the linearized theory. His consistency conditions were that the divergencelessness of the lowest-order equations needs to be enough to ensure the divergencelessness of the higher order terms entering in the right-hand side, to avoid the appearance of extra restrictions. For spin-1 fields, he showed that this was enough to fully characterize the set of consistent non-linear theories (Yang-Mills theories). However, in the case of Fierz-Pauli theory, he showed the existence of two families of theories: one corresponding to diffeomorphism-invariant theories and another one which was not diffeomorphism-invariant. This second family does not couple to its own energy-momentum tensor and, in that sense, cannot be obtained through the bootstrapping procedure {\it à la} Feynmann sketched above. In any case, he gave plausibility arguments for discarding the non-diffeomorphism-invariant theories: he argued that they would become inconsistent once one includes matter in the picture~\cite{Wald1986}. A related approach to that of Wald is the one taken by Ogievetsky and Polubarinov~\cite{Ogievetsky1965}, reaching also similar conclusions.

In more recent times, the work of Padmanabhan~\cite{Padmanabhan2008} questioning the uniqueness of Deser's construction attracted some interest back into the problem~\cite{Butcher2009,Deser2010}, and the issue of the uniqueness of the construction does not seem to be fully closed. To our knowledge there is no conclusive proof of the uniqueness of the construction yet, specially if one applies a strict bootstrapping procedure to all the fields of the linear theory. Indeed, Deser's analysis in the first order formalism includes an extra field in the linear theory that is not actually bootstrapped but used as an auxiliary field so that the procedure works straightforwardly. This extra auxiliary field ends up being the connection. Even if we ignore the presence of additional auxiliary fields, Deser's proof is of existence of a solution, and not of uniqueness of it. Nevertheless, analysis like that of Butcher \emph{et al.}~\cite{Butcher2009} do not seem to leave the room for other solutions to the self-coupling problem. Specially interesting are the recent works on the self-coupling of higher derivative theories \cite{Butcher2009,Deser2017,Ortin2017}. Whereas in~\cite{Butcher2009,Ortin2017} it was found that higher-derivatives theories bootstrap in the same sense than GR does, in~\cite{Deser2017}, by working in first-order formalism,  the author concluded that they do not unless one enforces ``by hand'' the condition that the connection is the Levi-Civita connection associated with the full metric.

The discussion outlined above is the departure point of our analysis here. In this work, we review different notions and key concepts related to the bootstrapping procedure, and extend the machinery developed by Butcher \emph{et al.} \cite{Butcher2009} to apply it to arbitrary metric-affine theories of gravity, namely, theories where the metric and the connection are treated as independent field variables.
Our aim with this paper is threefold. First, we aim to clarify certain points on the bootstrapping ongoing discussions, and offer original perspectives on the problem. Second, we intend to formulate metric-affine theories of gravity in a language that is closer to the one employed by particle physicists, namely, that of a field theory on top of Minkowski spacetime. Hopefully, such formulation is more amenable to extend the analysis of these theories to the quantum realm. Finally, we intend to clarify whether the presence of the new fields associated to the general connection (torsion and nonmetricity) spoil or modify somehow the bootstrapping of the metric sector, and whether they obey similar bootstrapping equations that couple them to other physically relevant sources.

The paper is structured as follows. In Section~\ref{Sec:Bootstrapping}, we give a sketch of the bootstrapping idea. First, in Subsection~\ref{Subsec:Conserved_Currents}, we give a review of the definitions of the energy-momentum tensor and the spin-density tensor: both the canonical ones and the ones obtained varying the action. In Subsection~\ref{Subsec:Bootstrapping}, we explain the self-coupling problem of gravity to its own energy-momentum tensor and discuss the general idea behind it. The next two sections are the core of the article. In Section~\ref{Sec:Bootstrapping_Metric}, we focus on the bootstrapping of theories with a dynamical metric. In section~\ref{Subsec:Metric_Vacuum}, we review the analysis done by Butcher \emph{et al.}~\cite{Butcher2009} for metric theories of gravity in vacuum. Then, in Subsection~\ref{Subsec:Metric_Matter}, we extend their analysis to consider arbitrary bosonic matter content and, in Subsection~\ref{Subsec:Metric_Torsion}, we include torsion in the discussion. In Section~\ref{Sec:Bootstrapping_Vielbein}, we move on to discuss gravitational theories written in terms of the vielbein instead of the metric. First, in Subsection~\ref{Subsec:Vielbein_Metric}, we show how previous results are recovered if the action depends on the vielbein only through the metric. Then we progress to include fermionic matter and torsion in Subsection~\ref{Subsec:Vielbein_Torsion}. Section~\ref{Sec:Nonmetricity} is devoted to the study of the influence of nonmetricity in the self-coupling problem. We have dedicated a separate section for it since nonmetricity is special in the sense that it couples to a source which does not always admit a definition as a canonical current. In Section~\ref{Sec:Unimodular}, we give a brief discussion of how our results can be easily embedded in unimodular gravity theories, where we simply need to replace the energy-momentum tensor by its traceless counterpart. Finally we finish in Section~\ref{Sec:Conclusions} by summarizing the work and discussing the conclusions that can be drawn up from our analysis as well as pointing future directions of work.

At the end, the reader can find two appendices. In Appendix~\ref{App:DecompositionConnection}, we reproduce some useful expressions regarding the decomposition of a general connection, and Appendix~\ref{App:General_Fields} is devoted to the derivation of some required identities. Now we proceed to collect our notation and conventions.

%-------------------------------------------------------
 \paragraph*{\textbf{Notation and conventions.}}
%-------------------------------------------------------

In this article, we use the signature $(-,+,...,+)$ for the spacetime metric and natural units  $c=\hbar=1$.  We also introduce $H_{(\mu\nu)}:= \frac{1}{2!}(H_{\mu\nu}+H_{\nu\mu})$ and $H_{[\mu\nu]}:= \frac{1}{2!}(H_{\mu\nu}-H_{\nu\mu})$, and similarly for an object with $n$ indices instead of 2. The symbol $\dimM$ represents the dimension of the spacetime manifold. Greek indices ($\mu,\nu,\rho,\lambda...$) refer to arbitrary coordinates in spacetime and run from $0$ to $\dimM-1$, lowercase Latin indices ($a,b,c...$) represent anholonomic indices in an arbitrary orthonormal frame, and uppercase Latin indices ($A,B,C...,I,J...$) are generic indices used to label the elements of a certain set of fields and their components. Einstein's summation convention is used throughout the work unless otherwise stated. We will use ``flat spacetime'' as a synonym of Minkowski spacetime (not to be confused with teleparallel spacetimes).

For the curvature tensors we use the conventions in the book of Wald~\cite{Wald1984}, {\it i.e.}, $[\nabla_\mu, \nabla_\nu] V^\lambda =: - R_{\mu\nu\rho}{}^\lambda V^\rho$, $R_{\mu\nu}:=R_{\mu\lambda\rho}{}^\lambda$. For the indices of the affine connection, we use the convention that can be read from the following expression for the covariant derivative: $\nabla_\mu V^\rho=\partial_\mu V^\rho + \Gamma_{\mu\nu}{}^{\rho}V^\nu$.

We use $g_{\mu\nu}$ for an arbitrary metric and $\bar{g}_{\mu\nu}$ for the background metric. Indeed, we generically use small bars over background quantities ($\bar{g},\bar{\Phi},\bar{K}...$).

Concerning the affine connection, the symbols ($\LCg$, $\CDg$, $\Rg$, $\mathring{\omega}$...) correspond to the Levi-Civita connection of $g_{\mu\nu}$, ($\bar{\Gamma}$, $\bar{\nabla}$, $\bar{R}$...) are referred to the Levi-Civita connection of the background metric $\bar{g}_{\mu\nu}$, ($\GCD$, $\hat{R}$, $\hat{\omega}$) correspond to a general metric-compatible (but torsionful) connection, and if nothing is indicated, ($\Gamma$, $\omega$), the connection is assumed to be fully general, with both torsion and nonmetricity. For the definitions of torsion, nonmetricity, contorsion and disformation that we are using, the reader can check Appendix~\ref{App:DecompositionConnection}.

Throughout this paper, we work with an arbitrary set of gamma matrices satisfying $\gamma^\mu\gamma^\nu+\gamma^\nu\gamma^\mu = -2 g^{\mu\nu}$.

We now clarify some notation regarding functional derivatives and evaluation of parameters. Consider a generic functional $S[\mathrm{Q}^I]$ depending on a certain family of fields that we denote collectively as $\{\mathrm{Q}^I (x) \}$. In this work, we will use the notation 
%-------------------------------------------------------
\begin{equation}
\frac{\delta}{\delta \bar{\mathrm{Q}}^{I_1} (x_1) }\ldots\frac{\delta}{\delta \bar{\mathrm{Q}}^{I_k} (x_k) } S[\bar{\mathrm{Q}}^{I} + \lambda \mathrm{q}^{I}] :=\frac{\delta^k S[\mathrm{Q}^{I}]}{\delta \mathrm{Q}^{I_1} (x_1)\ldots\delta \mathrm{Q}^{I_k} (x_k)}\Big|_{\mathrm{Q}^{I}=\bar{\mathrm{Q}}^{I} + \lambda \mathrm{q}^{I}}. \label{eq:deltaSbacknotation}
\end{equation}
%-------------------------------------------------------
Notice that if the background $\{\bar{\mathrm{Q}}^{I}\}$ is a generic one, the left-hand side can be read literally, {\it i.e.} as {\it first evaluating the action in $\bar{\mathrm{Q}}^{I} + \lambda \mathrm{q}^{I}$ and then varying with respect to $\bar{\mathrm{Q}}^{I}$}. Additionally, we will also use the shortcut:
%-------------------------------------------------------
\begin{align}
   \left[ \int \dd^\dimM x\  \mathrm{q}^I(x) \frac{\delta}{\delta \bar{\mathrm{Q}}^{I} (x) }\right]^n := \int \dd^\dimM x_1\  \mathrm{q}^{I_1}(x_1) \frac{\delta}{\delta \bar{\mathrm{Q}}^{I_{1}}(x_1) } \ldots \int \dd^\dimM x_n\  \mathrm{q}^{I_n}(x_n) \frac{\delta}{\delta \bar{\mathrm{Q}}^{I_{n}}(x_n) }\,.
   \label{Eq:Shorcut_Notation}
\end{align}
%-------------------------------------------------------
Regarding the evaluation of parameters, we represent
%-------------------------------------------------------
\begin{equation}
    ...= A(\lambda,...)|_{\lambda=0,...} := A(0,...),
\end{equation}
%-------------------------------------------------------
where $A$ can be a big expression. Namely, everything from the equal sign on should be computed before the evaluation. Otherwise, it will be explicitly indicated with a square bracket:
%-------------------------------------------------------
\begin{equation}
    ...= A(\lambda,...) [B(\lambda,...)]_{\lambda=0,...} := A(\lambda,...) [B(0,...)].
\end{equation}
%-------------------------------------------------------

%------------------------------------------------
\section{Coupling gravity: Conserved currents and bootstrapping}
\label{Sec:Bootstrapping}
%------------------------------------------------

In order to familiarize the reader with the bootstrapping procedure that we will later detail and apply to metric-affine theories, in this section we will provide a general overview of the key aspects of the process. 
%------------------------------------------------
\subsection{Coupling to conserved currents}
\label{Subsec:Conserved_Currents}
%------------------------------------------------

For gravitational theories, there is a preferred current to which metric and/or vielbein perturbations usually couple, both in purely metric and more general theories. This is the energy-momentum tensor. In the case that an independent connection is included in the theory, we expect torsion to couple to the spin-density current, and nonmetricity to the so-called dilation and shear currents, which are associated to ${\rm GL}(\dimM,\mathbb{R})$-transformations outside the Lorentz group~\cite{Hehl1995}. We will mainly focus on the canonical currents of the Poincaré group and comment on the aspects concerning the nonmetricity in Section~\ref{Sec:Nonmetricity}.

In the present section, we will provide the definition of both the energy-momentum tensor and the spin-density current through two different prescriptions. The canonical prescription defines these objects as Noether currents for a theory defined on flat spacetime, and the Einstein-Hilbert-like prescription defines them through a variational prescription. We will also discuss the presence of ambiguities in both prescriptions, showing how they lead to conserved currents generally differing by an identically divergenceless term. This means that the Noether charges associated to both of the currents are the same.

~

%------------------------------------------------
\paragraph*{\textbf{Canonical currents:}}
%------------------------------------------------

Let us begin with the canonical definition of spin-density current and energy-momentum tensor. Consider a field theory described by a Lagrangian of the type $\mathcal{L}=\mathcal{L}(\Phi^A,\partial_\mu\Phi^A)$ which is invariant under the Poincar\'e group. Then, the Noether currents associated with translations and Lorentz transformations, namely the energy-momentum and the angular momentum tensors, are respectively given by
%------------------------------------------------
\begin{align}
    T_\text{can\,}{}^\mu{}_\nu &:= \frac{\partial \mathcal{L}}{\partial \partial_\mu \Phi^A}\partial_\nu\Phi^A - \mathcal{L} \delta^\mu{}_{\nu}\,,\label{Eq:CanonicalTmunu}\\
    J_\text{can\,}{}^{\mu\nu\lambda} &:= T_\text{can\,}{}^{\mu [\lambda}x^{\nu]} + S_\text{can\,}{}^{\mu\nu\lambda}\,,\label{Eq:CanonicalAngularMom}
\end{align}
%------------------------------------------------
where we have introduced the canonical spin-density tensor
%------------------------------------------------
\begin{equation}
    S_\text{can\,}{}^{\mu\nu\lambda} := \frac{1}{2} \sum_A\frac{\partial \mathcal{L}}{\partial \partial_\mu\Phi^A} (\Lambda^{\nu\lambda}_{{\Phi}^A}) \Phi^A
\end{equation}
%------------------------------------------------
and where $\Lambda^{\nu\lambda}_{{\Phi}^A}$ is the generator of the Lorentz algebra that correspond to the representation under which the field $\Phi^A$ transforms.

For concreteness and later use, let us provide these currents explicitly for several familiar field theories in flat spacetime and in Cartesian coordinates. For the massless free scalar field, described by
%------------------------------------------------
\begin{align}
    S[\Phi] = - \frac{1}{2} \int \dd^\dimM x\ \eta^{\mu \nu} \partial_{\mu } \Phi \partial_{\nu} \Phi,
    \label{Eq:Scalar_Field_action}
\end{align}
%------------------------------------------------
we obtain the currents
%------------------------------------------------
\begin{align}
    & T_\text{can}{}_{\mu \nu} = -  \partial_{\mu} \Phi \partial_{\nu} \Phi + \frac{1}{2} \eta_{\mu \nu} \partial_{\rho } \Phi \partial^\rho \Phi,
    \label{Eq:Scalar_Field_EMT}\\
    & S_\text{can\,}{}^{\mu\nu\lambda} =0\,;
\end{align}
%------------------------------------------------
and for the massive Dirac spinor, described by
%-------------------------------------------------------
\begin{align}
    S[\Psi] = \int \dd^4 x \left[ \frac{\ii}{2} \big(\CPsi \gamma^\mu \partial_\mu \Psi- \partial_\mu\CPsi \gamma^\mu \Psi \big)  - m \CPsi \Psi  \right]\label{Eq:SDirac}
\end{align}
%------------------------------------------------
we obtain
%------------------------------------------------
\begin{align}
    T_\text{can\,}{}_{\mu\nu} &= \frac{\ii}{2}(\CPsi \gamma_\mu\partial_\nu\Psi -\partial_\nu \CPsi \gamma_\mu\Psi) -  \left[ \frac{\ii}{2} \big(\CPsi \gamma^\rho \partial_\rho \Psi- \partial_\rho \CPsi \gamma^\rho \Psi \big)  - m \CPsi \Psi  \right] \eta_{\mu\nu}\,,\label{Eq:EMTensorcanDirac}\\
    S_\text{can\,}{}^{\mu\nu\lambda} &=  \frac{\ii}{4} \CPsi \gamma^{[\mu}\gamma^\nu\gamma^{\lambda]}\Psi\,.\label{Eq:SpinTensorcanDirac}
\end{align}
%------------------------------------------------
To derive these expressions we have used $\Lambda^{\nu\lambda}_{\Phi}=0$ for the scalar, $\Lambda^{\nu\lambda}_{\Psi}=\frac{1}{2}\gamma^{[\nu}\gamma^{\lambda]}$,
 and $\Lambda^{\nu\lambda}_{\CPsi}=-\frac{1}{2}\gamma^{[\nu}\gamma^{\lambda]}$ for the spinor and its adjoint, as well as the following property of the Dirac matrices $\gamma^\mu \gamma^{[\nu}\gamma^{\lambda]}+ \gamma^{[\nu}\gamma^{\lambda]}\gamma^\mu = 2 \gamma^{[\mu}\gamma^\nu\gamma^{\lambda]}$.

The physical meaning of Noether currents stems from its conservation on-shell if the system satisfies the corresponding global symmetry. However, this conservation is not jeopardized by adding to the current terms that are identically divergenceless. As we will see, these terms play a determinant role in the bootstrapping procedure that we will outline. From the terms that can be added without spoiling conservation of the current, those which are identically conserved can be expressed as the divergences of certain objects usually called superpotentials. For instance, for the energy-momentum tensor current we can always add a term of the type $\Delta T^{\mu \nu} =  \partial_{\rho} \chi^{[ \rho \mu] \nu}$. To illustrate this, let us explicitly show an example that we will use later: for the case of the massless scalar field in Eq.~\eqref{Eq:Scalar_Field_action}, an example of such a term is
%------------------------------------------------
\begin{align}
    \Delta T_{\mu \nu} = \alpha \left( \partial_{\mu} \partial_{\nu} \Phi - \eta_{\mu \nu} \partial^2 \Phi \right),
    \label{Eq:Example_SPT}
\end{align}
%------------------------------------------------
where the associated superpotential is
%------------------------------------------------
\begin{align}
    \chi^{\rho \mu \nu} = 2  \alpha \partial^{[\rho} \Phi \eta^{\mu] \nu}
    \label{Eq:Example_SPT2}
\end{align}
%------------------------------------------------
and $\alpha$ is an arbitrary constant. In some cases, the canonical energy-momentum tensor \eqref{Eq:CanonicalTmunu} is not symmetric\footnote{Also, in some cases it is not a tensor, and it may not be gauge-invariant. It has been argued that the properly defined Noether current should be equivalent to the Hilbert energy-momentum tensor, and it is thus a somewhat misleading convention to call
(\ref{Eq:CanonicalTmunu}) ``the canonical energy-momentum tensor'' \cite{Gomes:2022vrc}. However, in this paper we stick to the conventional terminology.}, and it is possible to find an appropriate superpotential term to make it symmetric, as famously done in the Belinfante procedure~\cite{Belinfante1940,Ortin2017}. These same ambiguities would occur for the case of the full angular momentum tensor~\eqref{Eq:CanonicalAngularMom}.

~

%------------------------------------------------
\paragraph*{\textbf{Hilbert's prescription:}}
%------------------------------------------------

Let us now move on to the other method to compute the energy-momentum and spin-density tensor currents, known as Hilbert's prescription. Let us start by considering theories with the metric as a field variable. Note that in theories with spinor fields, the appropriate field variable would be the vielbein. We will analyze this case later.
In Hilbert's prescription, the starting point is a certain action defined in Minkoswki spacetime, and the prescription is implemented in three steps:
\begin{enumerate}
\item Firstly, we extend the action to a curved spacetime by following a minimal-like coupling procedure. If we are only interested in deriving the energy-momentum tensor, it is enough to promote the Minkowski metric to a general one $g_{\mu\nu}$, as well as $\partial_\mu$ to covariant derivatives with respect to the Levi-Civita connection of $g_{\mu\nu}$, which we denote as $\CDg$. This is the \emph{standard} Hilbert's prescription. However, since we are interested in obtaining the spin-density current in a similar way, this first step should be generalized as follows: we promote the metric to a general one and the derivatives $\partial_\mu$ to covariant derivatives with respect to a torsionful (but metric-compatible) connection. Notice that now the result is the same we obtained with the previous prescription, but with extra terms that go with the difference between the new connection and the Levi-Civita one, {\it i.e.}, the contorsion tensor $K_{\mu\nu}{}^\rho$ introduced in Eq.~\eqref{Eq:defContorsion}.

\item The resulting matter action should be understood as a functional of the metric, the contorsion and the matter fields, $S_{\text{M}}[g, K, \Phi]$. The second step is to vary such an action with respect to the metric and the contorsion field.

\item Finally, we evaluate the geometrical fields in the resulting expressions by setting a torsion-free Minkowski spacetime, which leads to the definitions
\end{enumerate}
%------------------------------------------------
\begin{align}
    T_{\text{H}}{}_{\mu \nu} &:= \frac{-2}{\sqrt{-g}} \frac{\delta S_{\text{M}}[g,K,\Phi]}{\delta g^{\mu \nu}} \bigg\rvert_{ g^{\mu \nu} = \eta^{\mu \nu}, K_{\mu\nu}{}^{\rho} = 0}, \label{Eq:defTH}\\
    S_{\text{H}}{}^{\mu \nu \lambda} \eta_{\lambda\rho} &:= \frac{1}{\sqrt{-g}} \frac{\delta S_{\text{M}}[g,K,\Phi]}{\delta K_{\mu\nu}{}^{\rho}} \bigg\rvert_{ g^{\mu \nu} = \eta^{\mu \nu}, K_{\mu\nu}{}^{\rho} = 0}. \label{Eq:defSH}
\end{align}
%------------------------------------------------
Note that, by construction, the tensors obtained from this procedure always fulfill the symmetries
%------------------------------------------------
\begin{equation}
    T_{\text{H}}{}_{\mu \nu}=T_{\text{H}}{}_{(\mu \nu)}\,,\qquad S_{\text{H}}{}^{\mu \nu \lambda}=S_{\text{H}}{}^{\mu [\nu \lambda]}\,.
\end{equation}
%------------------------------------------------
The latter is a consequence of the fact that the contorsion tensor is antisymmetric in the last two indices $K_{\mu\nu\rho}=K_{\mu[\nu\rho]}$. Notice that, to be precise, spin-density current is not a conserved current in relativistic field theories on top of Minkowski spacetime. It is part of the angular momentum current, which is conserved. However, this fact is not relevant for the bootstrapping of the contorsion.

In order to bring closer this procedure with the Noether procedure to derive canonical currents, note that there is a counterpart to the ambiguities that arose in such procedure. The possibility of adding an identically divergenceless term to the canonical currents now manifests in the possibility of adding non-minimal couplings in the action that identically vanish in flat spacetime. To see this more explicitly, let us revisit the case of the scalar field theory in Eq.~\eqref{Eq:Scalar_Field_action}. The action in curved spacetime is
%------------------------------------------------
\begin{align}
    S_{\text{M}}[\Phi] =  - \frac{1}{2} \int \dd^\dimM x\,\sqrt{-g}\  g^{\mu \nu} \partial_{\mu} \Phi \partial_{\nu} \Phi,
\end{align}
%------------------------------------------------
and it is not difficult to check that the corresponding Hilbert energy-momentum tensor coincides with the result in Eq.~\eqref{Eq:Scalar_Field_EMT}. However, one can add to the scalar field action a non-minimal coupling of the type
%------------------------------------------------
\begin{align}
    S_{\text{nm}}[g, \Phi] = - \frac{\alpha}{2} \int \dd^\dimM x\, \sqrt{-g}\ \Phi \Rg(g).
\end{align}
%------------------------------------------------
which yields a correction to the Hilbert energy-momentum tensor with exactly the same form as the one arising from the superpotential in Eq.~\eqref{Eq:Example_SPT2} in the canonical prescription.

~

%------------------------------------------------
\paragraph*{\textbf{Extension of the Hilbert prescription to the vielbein formulation:}}
%------------------------------------------------

For Dirac spinors, for example, the prescription described in the previous subsection does not work, since the promotion to curved spacetime requires extra structure (a frame field or vielbein). The idea now is to follow essentially the same steps as before, {\it i.e.}, we promote $\eta_{\mu\nu}\to g_{\mu\nu}$ and $\partial_\mu$ to a metric-compatible connection, and we also introduce the vielbein field as
%------------------------------------------------
\begin{equation}
    g^{\mu\nu} = \eta^{ab} e^\mu{}_a e^\nu{}_b,\label{Eq:gtoe}
\end{equation}
%------------------------------------------------
work with the spin connection $\omega_{\mu a}{}^b$ instead of $\Gamma_{\mu\nu}{}^\rho$ and promote the gamma matrices to curved space as $\gamma^\mu \to \gamma^\mu = \gamma^a e^\mu{}_a$. As a result we get a \emph{covariantized} matter action of the form $S_{\text{M}}[e,K,\Phi]$, where $K$ now refers to the contorsion with two anholonomic indices
%------------------------------------------------
\begin{equation}
    K_{\mu a b} := K_{\mu\nu}{}^\rho e^\nu{}_a e_\rho{}^b \eta_{cb} \,, \label{Eq:defKmuab}
\end{equation}
%------------------------------------------------
which is nothing but the difference between the torsionful and the Levi-Civita spin connections. The next steps are totally analogous to the ones described in the previous subsection: we vary with respect to $e^\mu{}_a$ and $K_{\mu a b}$ and evaluate in torsionfree Minkowski spacetime:
%------------------------------------------------
\begin{align}
    T_{\text{V}\,}{}_{\mu\nu}{} &:= e_\nu{}^c\eta_{ca}\frac{1}{|e|} \frac{\delta S_{\text{M}}[e,K,\Phi]}{\delta e^\mu{}_a}\bigg\rvert_{ e^{\mu}{}_a = \delta^\mu{}_a, K_{\mu ab} = 0}\,, \label{Eq:defTV}\\
    S_{\text{V}\,}{}^{\mu \nu\lambda}&:=e^\nu{}_a e^\lambda{}_b\frac{1}{|e|} \frac{\delta S_{\text{M}}[e,K,\Phi]}{\delta K_{\mu ab}}\bigg\rvert_{ e^{\mu}{}_a = \delta^\mu{}_a, K_{\mu ab} = 0}\,.\label{Eq:defSV}
\end{align}
%------------------------------------------------
At this point, it is important to notice that for the theories described in the previous subsection ({\it i.e.}, those that do not require a vielbein field) the tensors in Eqs.~\eqref{Eq:defTV}-\eqref{Eq:defSV} essentially coincide with the ones in Eqs.~\eqref{Eq:defTH}-\eqref{Eq:defSH}. This can be easily seen by performing a chain rule, taking into account Eqs.\eqref{Eq:gtoe} and \eqref{Eq:defKmuab}. Therefore, we have just generalized the previous prescription to a more general class of matter fields. For instance, for the Dirac Lagrangian in Eq.~\eqref{Eq:SDirac} one gets that the Hilbert energy-momentum and spin-density coincide with the canonical ones (see Eqs.~\eqref{Eq:EMTensorcanDirac}-\eqref{Eq:SpinTensorcanDirac}),
\begin{equation}
    T_{\text{V}\,}{}_{\mu\nu} = T_{\text{can}\,}{}_{\mu\nu}\,,\qquad S_{\text{V}\,}{}^{\mu \nu\lambda} = S_{\text{can}\,}{}^{\mu \nu\lambda}\,.
\end{equation}
%------------------------------------------------

%------------------------------------------------
\subsection{Bootstrapping: An overview}
\label{Subsec:Bootstrapping}
%------------------------------------------------

Let us now briefly outline the idea of the bootstrapping that we will have in mind through the rest of the manuscript. Consider a linear theory with a given number of degrees of freedom. Our aim is to find a non-linear theory with the same number of degrees of freedom that reduces to the previous one in its linear regime. This might be trivial in some cases. However, when the linear theory satisfies some gauge symmetry, the self-couplings that one can add that preserve the number of degrees of freedom are very limited. Generically, one looks for an iterative prescription such that, starting from a quadratic action, self-couplings which are consistent with the gauge symmetry are generated order by order. This is done by coupling the theory to a conserved current that needs to be identified within the set of conserved currents from the free theory. In this way, the process leads to a self-interacting non-linear theory that satisfies a non-linearly deformed version of the linear gauge symmetry. To successfully implement this program, we need to identify a suitable conserved current in the linear theory. Here we want to remark that, if two different superpotentials are chosen for the linear theory (which does not change the physical content of the currents), the resulting iterative procedure will in general yield different non-linear theories, in which the non-linear extensions of the symmetries (if they exist) will differ.

At the practical level, a way to implement the procedure is to find currents that satisfy order-by-order conservation. The introduction of the self-coupling to the current at a given order will in general induce non-trivial terms in it at the next order, which will demand for further higher-order self-couplings. The process will thus finish if the self-coupling does not produce higher-order terms for the current at a given order. Otherwise, the procedure will extend up to infinite order and, to be successful, we would need to find a general relation between the theory to a given arbitrary order and the currents of the next order.

To be more specific, assume that we have a set of fields which we denote as $\Phi^A$, being $A$ a collective index including possible internal indices or spacetime indices. For an action that is quadratic on the fields, we will find linear equations of motion of the form
%------------------------------------------------
\begin{align}
    \mathcal{D}_{A B} \Phi^B = 0,
\end{align}
%------------------------------------------------
where $\mathcal{D}_{A B}$ is a given differential operator. Assume that we identify a conserved current $j^A$ which we want to add as a source, so that it couples as
%------------------------------------------------
\begin{align}
    \mathcal{D}_{A B} \Phi^B = \lambda  j_{A},
\end{align}
%------------------------------------------------
where the coupling $\lambda$ will be useful as a bookkeeping dimensionless parameter for the iterative procedure. In order to derive this coupling from a variational principle, we would need to add a term to the action which would schematically be of the form
%------------------------------------------------
\begin{align}
    \Delta S \sim  \lambda \int \dd^\dimM x\  j_A \Phi^A.
\end{align}
%------------------------------------------------
However, because $j_A$ will generally have a nontrivial dependence on $\Phi^A$, the term $\Delta S$ will also contribute to the right hand side of the field equations, so that the source will now get an additional contribution $j^A+\lambda \Delta j_A $. This means that, if we want an action to derive the corrected equations, we would need to add an additional term of order $\lambda^2$ to the action. This situation will repeat yielding an iterative process. Indeed, one sufficient condition for the process to finish in a finite number of steps $N$, is that the order $\lambda^N$ term that we add to the current does not contain derivatives of the fields, so that no $\lambda^{N+1}\Delta j_A$ has to be added at that order. Indeed, from the form of Noether currents, it is clear that if the $\Delta j_A$ at order $N$ does not contain derivatives of the fields it will not contribute to these currents. This is what occurs already at order $N=2$ if one implements this procedure for the gauge invariant spin-1 field~\cite{Deser1970}. In other cases the recursive process might need of an infinite number of steps, as it occurs for the case of the spin-2 field theory~\cite{Butcher2009}. The inclusion of additional fields into the picture can be done in a straightforward manner: the iterative procedure will be virtually the same but now the initial seed would be a conserved current including also those additional fields what we need to use as a source.

Let us now dive into the gravitational case and its subtleties. The fields of the theory will be the inverse metric perturbations, which we call $h^{\mu \nu}$. In the case of quadratic equations, there are only two possible maximum set of gauge symmetries allowed to get rid of all the unphysical degrees of freedom propagated by a symmetric rank-2 tensor beyond the two standard gravitational polarizations (which correspond to a massless spin-2 field excitation if the background satisfies a Poincar\'e symmetry). These are linearized diffeomorphisms, which lead to the Fierz-Pauli action, or the group of Weyl transformations and Transverse Diffemorphisms (WTDiff group), which are a linearized version of Unimodular Gravity~\cite{Alvarez2006,Carballo2022}. Let us focus on the case in which the theory is invariant under linearized diffeomorphism, which will be the main focus of this work, although we will also comment on theories which are invariant under the WTDiff symmetries in Section~\ref{Sec:Unimodular}.

For the case of linearized diffeomorphism-invariant theories, we will have an equation of the form
%------------------------------------------------
\begin{align}
    \mathcal{D}_{\mu \nu \rho \sigma} h^{\rho \sigma } = 0,
    \label{eq:schematiclinearized}
\end{align}
%------------------------------------------------
with $\mathcal{D}_{\mu \nu \rho \sigma}$ a suitable differential operator. These equations are such that they are divergenceless and symmetric in the indices $\mu$ and $\nu$, so that we need to couple them to an object which is symmetric and divergenceless on-shell. There is only one natural candidate which is the energy-momentum tensor, as it is the only object that we can construct with two indices which is symmetric and conserved on-shell. Hence we need to add to the right hand-side of Eq.~\eqref{eq:schematiclinearized} a term which would correspond to the ``gravitational energy-momentum tensor'', computed following any of the procedures outlined in the subsection above. %However, we know that there is no local definition of energy-momentum tensor for the gravitational field in a diffeomorphism-invariant theory. 
Such a tensor can be consistently defined in the context of the geometrical trinity \cite{BeltranJimenez2019,Gomes:2022vrc} which is outside the scope of this article, but otherwise it is known that there is no local definition of energy-momentum tensor for the gravitational field in a diffeomorphism-invariant theory. 
Therefore, either we construct a non-gauge-invariant tensor (which would be hard to be given a clear physical meaning), or we construct a non-conserved quantity~\cite{Weinberg1980,Barcelo2005,Barcelo2021b}. We will refer to the gravitational energy-momentum as the conserved tensor computed following the procedures outlined in the section above, and we will analyze the gauge invariance of the full non-linear action once the bootstrapping is finished. Adding the energy-momentum tensor of the matter fields will then be a straightforward procedure.

Once this gravitational energy-momentum tensor has been added to the right-hand side of Eq.~\eqref{eq:schematiclinearized}, we need to add a piece to the original action to derive the coupling from an action. This is how the bootstrapping procedure begins in this case with the subtlety that, unlike the case for the gauge invariant spin-1 field, the iterative process is infinite. Instead of following the recursive procedure described above, it is more instructive to follow another line of thought. From this other perspective, we will instead start with a non-linear theory with a given set of properties and then analyze carefully whether it could be reconstructed from its linearization in an iterative way, in the line of \cite{Butcher2009}. The main aim of this work is to apply these ideas to the framework of metric-affine theories of gravity and clarify some issues that are relevant in its application to GR.

%------------------------------------------------
\section{Bootstrapping theories with a metric}
\label{Sec:Bootstrapping_Metric}
%------------------------------------------------

In this section we will focus on the bootstrapping for purely metric theories of gravity, {\it i.e.}, theories in which the gravitational sector is described only by a metric. We will put special emphasis on the relevance of the superpotentials chosen for the current and try to clarify their role for the success of the iterative procedure. 

%------------------------------------------------
\subsection{Gravity in vacuum}
\label{Subsec:Metric_Vacuum}
%------------------------------------------------

%In this section we will closely follow the discussion from Butcher \emph{et al.}~\cite{Butcher2009}. Let us consider an arbitrary metric theory of gravity described by a diffeomorphism-invariant action $S[g]$, with $g$ representing the metric. We can perform an expansion of the form

In this section we will closely follow the discussion from Butcher \emph{et al.}~\cite{Butcher2009}. Our aim is to construct an arbitrary metric theory of gravity described by a diffeomorphism-invariant action $S[g]$, with $g$ representing the metric, starting from the action of a free field $h$ on a fixed metric background, such as the Minkowski metric $\eta$. Assuming that this action is quadratic in $h$, we will denote it by $S^{(2)}[\eta, h]$. To motivate this approach, let us assume that we have already found the full action $S[g]$. For the metric, we can perform an expansion of the form
%------------------------------------------------
\begin{align}
    g^{\mu\nu} = \bar{g}^{\mu\nu} + \lambda h^{\mu\nu},
\end{align}
%------------------------------------------------
where we are choosing $\bar{g}^{\mu\nu}$ to be a \textit{generic} solution to the vacuum field equations, {\it i.e.},
%------------------------------------------------
\begin{align}
    \frac{\delta S [g]}{\delta g^{\mu\nu}} \bigg\rvert_{g^{\mu\nu} = \bar{g}^{\mu\nu}} = 0,
    \label{Eq:Background}
\end{align}
%------------------------------------------------
and $h^{\mu\nu}$ represents the deviation with respect to the non-dynamical background $\bar{g}^{\mu\nu}$. Explicitly we have that the expansion is given by
%------------------------------------------------
\begin{align}
    S[g] = \sum_{n = 0}^{\infty} \lambda^n S^{(n)} [\bar{g}, h],
\end{align}
%------------------------------------------------
with the partial actions reading
%------------------------------------------------
\begin{align}
    S^{(n)} [\bar{g}, h] = \frac{1}{n!} \frac{\dd^n}{\dd \lambda^n} S [\bar{g} + \lambda h] \bigg\rvert_{\lambda  = 0}.
    \label{Eq:Partial_Actions}
\end{align}
%------------------------------------------------
Note that the $n$-th partial action contains $n$ powers of the fields. Given a field configuration for $\bar{g}^{\mu\nu}$ and $h^{\mu\nu}$, the derivatives with respect to $\lambda$ are just total derivatives for a function of a real variable. However, they can be written as functional derivatives with respect to the background metric by means of the relation
%------------------------------------------------
\begin{align}
    \frac{\dd}{\dd \lambda} S[\bar{g} + \lambda h] = \int \dd^\dimM x \ h^{\mu\nu} (x) \frac{\delta}{\delta \bar{g}^{\mu\nu} (x) } S[\bar{g} + \lambda h],
    \label{eq:LambdaDerivativeAsFunctionalDerivative}
\end{align}
%------------------------------------------------
where we are using the notation~\eqref{eq:deltaSbacknotation}. This expression holds up to surface integrals that arise when integrating by parts. These terms are neglected since they do not contribute to the equations of motion and do not give rise to any contribution to the energy-momentum tensor as well. From now on, we will simply skip them. 

By applying repeated differentiation, we can express $n$-th $\lambda$-derivatives as
%------------------------------------------------
\begin{align}
    \frac{\dd^n}{\dd \lambda^n} S[\bar{g} + \lambda h] = \left[ \int \dd^\dimM x\ h^{\mu\nu}(x) \frac{\delta}{\delta \bar{g}^{\mu\nu} (x) }\right]^n S[\bar{g} + \lambda h],
    \label{Eq:Partial_Derivatives}
\end{align}
%------------------------------------------------
where we are using the abbreviation \eqref{Eq:Shorcut_Notation}.
%------------------------------------------------
Therefore, we can combine Eqs.~\eqref{Eq:Partial_Actions} and~\eqref{Eq:Partial_Derivatives} to express all the higher-order partial actions $S^{(n)}$, for $n>2$, in terms of derivatives of $S^{(2)}[\bar{g}, h]$ as
%-------------------------------------------------------
\begin{align}
    S^{(n)}[\bar{g}, h] =  \frac{2}{n!} \left[ \int \dd^\dimM x\ h^{\mu\nu}(x) \frac{\delta}{\delta \bar{g}^{\mu\nu} (x) }\right]^{n-2} S^{(2)}[\bar{g}, h]. \label{eq:SnfromS2}
\end{align}
%-------------------------------------------------------
More details on the derivation of this equation can be found in Appendix~\ref{App:General_Fields_genformula}.

Note that, from the bottom-up approach, {\it i.e.}, starting from the linear theory in Minkowski spacetime, we do not know the full functional form of $S^{(2)}[\bar{g}, h]$ for any arbitrary background metric $\bar{g}^{\mu\nu}$, but only $S^{(2)}[\eta, h]$ evaluated at the fixed background metric $\eta^{\mu\nu}$, which thus lacks, for example, curvature or other non-minimal coupling terms that vanish at this flat background. So, in order to use the generating formula \eqref{eq:SnfromS2}, we need to promote  $\eta^{\mu\nu}\to\bar{g}^{\mu\nu}$ and add the appropriate non-minimal couplings. As shown in \cite{Butcher2009}, such a choice is crucial for the success of the bootstrapping procedure. Therefore, since $S^{(2)}[\bar{g},h]$ is all we need to build  the whole action for $h^{\mu\nu}$, the knowledge of the appropriate non-minimal couplings is equivalent to the knowledge of the full action. The fact that we can reconstruct all the partial actions from the quadratic piece of the action will allow us to show that the metric perturbations couple order by order to their own stress-energy tensor, namely that at each order in the expansion $S^{(n)}$, $h^{\mu\nu}$ couples to the energy-momentum tensor derived from the $S^{(n-1)}$ term. Hence the end product of the bootstrapping is a diffeomorphism-invariant theory provided that the right superpotential/non-minimal couplings are added order by order. To show that this holds explicitly, note that to lowest order in the expansion the equations of motion for the metric perturbation are
%-------------------------------------------------------
\begin{align}
    \lambda^2  \frac{1}{\sqrt{-\bar{g}}} \frac{\delta S^{(2)} [\bar{g}, h]}{\delta h^{\mu\nu}} = \order{\lambda^3}.
    \label{Eq:Lowest_Order_Bootstrapping}
\end{align}
%-------------------------------------------------------
In the case that $S[g]$ is the Einstein-Hilbert action, this would simply yield the Fierz-Pauli equation with non-minimal coupling terms that arise when linearizing the action around arbitrary backgrounds~\cite{Butcher2009}. We will come back to this example later. Now, we want to couple $h^{\mu\nu}$ to its own energy-momentum tensor. For such purpose, we can apply Hilbert's prescription, according to which the energy-momentum tensor associated with $S^{(2)}[\bar{g}, h]$ is
%-------------------------------------------------------
\begin{align}
    t^{(2)}_{\mu\nu} := - \frac{\lambda^2}{\sqrt{-\bar{g}}} \frac{\delta S^{(2)} [\bar{g}, h]}{\delta \bar{g}^{\mu\nu}}.
\end{align}
%-------------------------------------------------------
Strictly speaking, it would only correspond to the energy-momentum tensor that we have defined in Section~\ref{Sec:Bootstrapping} once we evaluate on $\bar{g}^{\mu\nu} = \eta^{\mu\nu}$ and up to a factor $2$, which we will absorb in the energy-momentum tensor definition for convenience. To build all the partial actions, we need to maintain the background metric $\bar{g}^{\mu\nu}$ arbitrary, {\it i.e.}, we need to know the partial actions $S^{(n)}[\bar{g},h]$ in an open neighbourhood of the flat spacetime metric $\eta^{\mu\nu}$. We want this object to appear to the next order on the right-hand side of Eq.~\eqref{Eq:Lowest_Order_Bootstrapping}, {\it i.e.}, we want the terms $\order{\lambda^3}$ to be precisely $\lambda t^{(2)}_{\mu\nu}$. Now, we need this term to be derivable from an action upon variation of $h^{\mu\nu}$, {\it i.e.}, we need a partial action $S^{(3)} [\bar{g}, h]$ such that
%-------------------------------------------------------
\begin{align}
     \frac{\lambda^3}{\sqrt{-\bar{g}}} \frac{\delta S^{(3)} [\bar{g}, h]}{\delta h^{\mu\nu}} = \lambda t^{(2)}_{\mu\nu},
\end{align}
%-------------------------------------------------------
so that $\lambda t_{\mu\nu}^{(2)}$ acts as source for the quadratic field equations of $h^{\mu\nu}$. Doing this to all orders would ensure that the field $h^{\mu\nu}$ couples to its own energy-momentum tensor order by order. This would be fulfilled trivially if we manage to show that the following equation holds at all orders in the perturbative expansion
%-------------------------------------------------------
\begin{align}
    \frac{\lambda^n}{\sqrt{- \bar{g}}}\frac{\delta S^{(n)} [\bar{g}, h]}{\delta h^{\mu\nu}} = \lambda t^{(n-1)}_{\mu\nu}.
    \label{Eq:Bootstrap_Eq_EMT}
\end{align}
%-------------------------------------------------------
Given the definition of the energy-momentum tensor through Hilbert's prescription, we have that the $n$-th order partial action contributes to the energy-momentum tensor as
%-------------------------------------------------------
\begin{align}
    t^{(n)}_{\mu\nu} := - \frac{\lambda^{n}}{\sqrt{-\bar{g}}} \frac{\delta S^{(n)} [\bar{g}, h]}{\delta \bar{g}^{\mu\nu}},
    \label{Eq:Gravitational_EMTensor}
\end{align}
%--------------------------------------------------------
we have that Eq.~\eqref{Eq:Bootstrap_Eq_EMT} can be rewritten as the key expression
%--------------------------------------------------------
\begin{align}
    \frac{\delta S^{(n)} [\bar{g},h]}{\delta h^{\mu\nu}} = \frac{\delta S^{(n-1)} [\bar{g},h]}{\delta \bar{g}^{\mu\nu}}.
    \label{Eq:Key_Bootstrap}
\end{align}
%--------------------------------------------------------
That this holds for arbitrary metric theories is proved in Appendix~\ref{App:General_Fields_funidentity}, see Eq.~\eqref{Eq:Key_Bootstrap_Gen}. Furthermore, since the action is diffeomorphism-invariant, the discussion in Subsection~\ref{Subsec:Conserved_Currents} guarantees that the right-hand side is precisely what we have been calling the energy-momentum tensor.

~

%------------------------------------------------
\paragraph*{\textbf{Example: General Relativity.}}
%------------------------------------------------

Let us now work out a simple example by applying this procedure to the linearized version of General Relativity. The Einstein-Hilbert action is given by
%------------------------------------------------
\begin{align}
    S[g] = \frac{1}{2 \kappa_{(\dimM)}}\ \int \dd^\dimM x \sqrt{-g}\, \Rg (g),
\end{align}
%------------------------------------------------
where $\Rg(g)$ represents the Ricci scalar tensor of the metric $g_{\mu\nu}$ and $\kappa_{(\dimM)}$ is the corresponding Einstein constant of the chosen dimension. Now we would need to expand the action in terms of a general metric $\bar{g}^{\mu \nu}$ and its perturbations on top of it, $h^{\mu \nu}$. This would result in a structure for the action which would be
%------------------------------------------------
\begin{align}
    S[\bar{g} + \lambda h] = \frac{1}{4 \kappa_{(\dimM)}} \int \dd^\dimM x \sqrt{- \bar{g}} &\sum_{n=2}^{\infty} \Big[  M_{(n)}{}^{\alpha_1 \alpha_2}{}_{\mu_1 \nu_1 \ldots \mu_n \nu_n } \bar{\nabla}_{\alpha_1 } h^{\mu_1 \nu_1} \bar{\nabla}_{\alpha_2}  h^{\mu_2 \nu_2}  h^{\mu_{3} \nu_{3} } \ldots h^{\mu_{n} \nu_{n}} \nonumber \\
    &\qquad  + H_{(n) \mu_1 \nu_1 \ldots \mu_n \nu_n} h^{\mu_1 \nu_1 } \ldots h^{\mu_n \nu_n}\Big],
    \label{Eq:SEH_MH}
\end{align}
%------------------------------------------------
where the tensors $M_{(n)}$ and $H_{(n)}$ are built only from curvature invariants, their covariant derivatives, the background metric tensor $\bar{g}_{\mu\nu}$, its inverse $\bar{g}^{\mu\nu}$, and the Kronecker $\delta^{\mu}{}_{\nu}$ tensor. This clearly has the structure of Eq.~\eqref{Eq:Partial_Derivatives}. As we have discussed and show explicitly in Appendix~\ref{App:General_Fields_genformula}, only the first term in this series is required to reconstruct the whole action, for which we have:
%------------------------------------------------
\begin{align}
    M_{(2)}{}^{\alpha \beta}{}_{\mu\nu \rho \lambda} &= -\frac{1}{2} \left[ \bar{g}^{\alpha\beta} \bar{g}_{ \mu (\rho} \bar{g}_{\lambda) \nu} - \bar{g}^{\alpha\beta}\bar{g}_{\mu\nu} \bar{g}_{\rho\lambda} - 2 \delta^{\alpha}{}_{(\rho} \bar{g}_{\lambda) (\mu} \delta^{\beta}{}_{\nu)} + \delta^{\alpha}{}_{(\rho} \delta^{\beta}{}_{\lambda)} \bar{g}_{\mu \nu} + \delta^{\beta}{}_{(\mu} \delta^{\alpha}{}_{\nu)} \bar{g}_{\rho\lambda} \right]\,, \nonumber \\
    H_{(2) \mu \nu \rho \lambda}  &=  \frac{1}{2} \bar{R} \left( \bar{g}_{\mu \rho} \bar{g}_{\lambda \nu} + \frac{1}{2} \bar{g}_{\mu \nu} \bar{g}_{\rho \lambda} \right) - \bar{R}_{\mu\nu} \bar{g}_{\rho\lambda}
\end{align}
%------------------------------------------------
(note that $M_{(2)}{}^{\alpha \beta}{}_{\mu\nu \rho \lambda}$ yields the Fierz-Pauli term in an arbitrary coordinate system). Although $H_{(n) \mu_1 \nu_1 \ldots \mu_n \nu_n}$ is identically zero when evaluated on the solutions to the vacuum equations of motion ({\it i.e.}, Ricci-flat backgrounds), it is needed for the self-coupling procedure to yield a coupling between $h^{\mu\nu}$ and its own stress-energy tensor. This is an instance of the arbitrariness that appears when writing the flat spacetime action on an arbitrarily curved spacetime, as it is explained in Subsection~\ref{Subsec:Conserved_Currents}. These terms are, as we explained before, irrelevant for the computation of Noether charges.

A comment about gauge invariance is in order now. The action that we have written is a version of GR, therefore, it will inherit diffeomorphism invariance. Let us assume that we truncate the action up to a given order $N$. This can be understood as the $(N-1)$-th order approximation to the full theory, where the non-linearities of GR of order $\order{\lambda^N}$ or higher are neglected. Therefore, only diffeomorphisms of $\order{\lambda^{N}}$ will be symmetries of this action. Namely, for a diffeomorphism generated by a given vector field $\xi^{\mu}$, it is necessary that the vector is $\mathcal{O}(1)$ in $\lambda$ if one wants to implement the transformation as a power series in $\lambda$ in a simple way. With this choice, the gauge transformations act on the field $h^{\mu \nu}$ simply by the combination of the action of the Lie derivative on the background metric and on $h^{\mu \nu}$ itself~\cite{Butcher2009} 
%------------------------------------------------
\begin{align}
     h^{\mu \nu} &\to h^{\mu \nu} + \delta h^{\mu \nu},  \nonumber \\
    \delta h^{\mu \nu} &= \frac{1}{\lambda} \sum_{n=1}^{N} \frac{\left( \lambda \mathcal{L}_{\xi} \right)^n}{n!} \bar{g}^{\mu \nu} + \sum_{n=1}^{N-1} \frac{\left( \lambda \mathcal{L}_{\xi} \right)^n}{n!} h^{\mu \nu}, 
\end{align}
%------------------------------------------------
where we have neglected $\order{\lambda^N}$ to be consistent with the assumed level of approximation. From the above action of gauge transformations in the truncated actions we can derive their action on any object built  with these tensors, e.g. the energy-momentum tensors $t^{(n)}_{\mu \nu}$ introduced above. 

Let us now discuss the problem raised by Padmanabhan in~\cite{Padmanabhan2008} regarding the statement that bootstrapping of the Fierz-Pauli action actually yields GR. In that reference, he argued that if one takes the Fierz-Pauli action in flat spacetime and replaces the flat metric and its Levi-Civita compatible derivatives with those of an arbitrary background, one is not able to reconstruct the whole Einstein-Hilbert action. This was also mentioned in~\cite{Butcher2009} and interpreted as saying that it is not possible to recover GR from the linear Fierz-Pauli theory through a bootstrapping procedure. However, we do think that this conclusion is reached due to them having a very strict definition of the energy-momentum tensor. Such definition does not allow the addition of superpotentials that do not contribute to the Noether charges in flat spacetime and, therefore, should be mild from the point of view of the Noether current. Indeed, note that the bootstrapping procedure can be applied blindly to any action $S^{(2)}[\bar{g},h]$, quadratic in $h^{\mu\nu}$. Given such an action, we can generate a whole family of $S^{(n)}[\bar{g},h]$ for $n>2$ by applying equation~\eqref{eq:SnfromS2} which will automatically verify  the bootstrapping equation \eqref{Eq:Key_Bootstrap}. For instance, this would happen for the action considered by Padmanabhan~\cite{Padmanabhan2008}. The whole point is that it might not be possible to identify at the end of the day that the resulting action is such that the background $\bar{g}^{\mu \nu}$ can be reabsorbed into a new metric degree of freedom $g^{\mu \nu} = \bar{g}^{\mu \nu} + h^{\mu \nu}$  (possibly up to boundary terms) yielding a diffeomorphism-invariant action. Consequently, we would end up with a background dependent theory in which the recursive relation~\eqref{Eq:Key_Bootstrap} does no longer mean that $h^{\mu\nu}$ couples to its own energy-momentum tensor. For actions that are not background-independent, Hilbert's prescription to compute the energy-momentum tensor is not well-defined due to the presence of two different metrics.  

In order for the bootstrapping to work as expected, the addition of the correct superpotential, or accordingly the correct non-minimal coupling, proves to be crucial. Indeed, if we start on applying the bootstrapping procedure to a linear theory of a massless spin 2 field, we only recover GR if we add the non-minimal coupling $H_{(2)}$ as well as the rest of the non-minimal couplings arising at each order in the process $H_{(n)}$. Otherwise we are not granted to find a diffeomorphism-invariant theory which couples to its energy-momentum. We can ask then the relation between these non-minimal couplings and the celebrated theorem by Lovelock~\cite{Lovelock1971}. This theorem states that in 4 dimensions there is only one diffeomorphism-invariant local action (up to boundary terms) yielding 2nd order field equations for a rank-2 symmetric and invertible tensor. However, starting from Fierz-Pauli action, one could virtually choose many different superpotentials to add to the partial energy-momentum tensors $t^{(n)}_{\mu\nu}$ (or non-minimal couplings to the partial actions $S^{(n)}$). Lovelock theorem then points to the existence of a class of superpotentials/non-minimal couplings that will yield GR (up to boundary terms) after the bootstrapping. We could thus wonder whether uniqueness in Lovelock's sense also implies uniqueness of the possible superpotentials/non-minimal terms that will lead to GR.  In light of what we have exposed before, it appears to us that the $H_{(n)}$ terms are indeed the only possible choice to recover GR through the bootstrapping procedure. This has the implication that the bootstrapped Fierz-Pauli with wrong choice of superpotential will necessarily spoil one of the axioms of Lovelock's theory, namely, the resulting theory will not be diffeomorphism-invariant. On the other hand, the uniqueness on the choice of superpotential points in the direction that there can be an alternative statement of Lovelock theorem in terms of superpotentials and the bootstrapping prescription. We leave for future work going beyond 4 dimensions, where we expect to be able to characterize how different choices of superpotentials/non-minimal couplings for the Fierz Pauli theory would lead us to the different Lovelock theories through the bootstrapping procedure.

%------------------------------------------------
\subsection{Gravity in the presence of arbitrary bosonic matter}
\label{Subsec:Metric_Matter}
%------------------------------------------------

In Butcher \emph{et al.}~\cite{Butcher2009}, they bring free matter fields into the picture. Let us now extend their results by considering an arbitrary matter content. Thus, we now consider an action $S[g,\Phi]$ that includes, in addition to the metric, some matter fields $\{\Phi^A\}$, where $A$ is a label covering all the matter fields as well as their internal and/or spacetime indices. We want again to expand around a given background for the matter and metric as
%------------------------------------------------
\begin{align}
    & g^{\mu\nu} = \bar{g}^{\mu\nu} + \lambda h^{\mu\nu}, \\
    & \Phi^A = \bar{\Phi}^{A} + \lambda \phi^A,
\end{align}
%------------------------------------------------
where in principle we can choose any solution to the equations of motions, {\it i.e.} $\bar{g}^{\mu\nu}$ and $\bar{\Phi}^{A}$ are such that
%--------------------------------------------------------
\begin{align}
    & \frac{\delta S[g, \Phi]}{\delta g^{\mu\nu}} \bigg\rvert_{g^{\mu\nu} = \bar{g}^{\mu\nu}} = 0\,, \\
   & \frac{\delta S[g, \Phi]}{\delta \Phi^A} \bigg\rvert_{ \Phi^A = \bar{\Phi}^A} = 0 \,.
\end{align}
%--------------------------------------------------------
Because we are mostly interested in bootstrapping from flat backgrounds, we will only consider vanishing matter backgrounds, as non-vanishing matter backgrounds are not consistent with a flat metric. Hence, we will from now on focus on the backgrounds with $\bar{\Phi}^{A}$ = 0 in the same vein that we specialize to the flat background $\bar{g}^{\mu \nu} = \eta^{\mu \nu}$. However, in order to be able to derive the stress energy-tensor with the Hilbert prescription we need to know the action in a neighbourhood of $\eta^{\mu\nu}$ to be able to compute variations. 

At this point, it is important to insist on this distinction between the fields associated with currents (the metric and, in subsequent sections, the contorsion, the vielbein and the nonmetricity) and the matter fields. In this paper, we focus on the generation of the non-linear equations for the former. Therefore we have to keep the background arbitrary to perform variations, and such variations commute with the evaluation of the background matter fields. The choice $\bar{\Phi}=0$ also allows us to follow closely the approach by Butcher \emph{et al.} \cite{Butcher2009}. In fact, the self-coupling of matter fields to their energy-momentum tensor has already been considered in the literature (see {\it e.g.} \cite{BeltranCembranos2018}). 

We will further assume that the action can be decomposed as a sector that simply contains $g^{\mu\nu}$ and a sector that contains both $g^{\mu\nu}$ and $\Phi^{A}$, usually called gravitational and matter sectors, so that
%--------------------------------------------------------
\begin{align}
    S[g,\Phi] = S_{\text{g}} [g] + S_{\text{M}} [g, \Phi].
\end{align}
%--------------------------------------------------------
We will also assume that the matter action can be written as
%--------------------------------------------------------
\begin{align}
     S_{\text{M}} [g, \Phi] &= \sum_{p=2}^{N}  \mathcal{A}_{\text{M}} ^{(p)} [g, \Phi], \\
     \mathcal{A}_{\text{M}} ^{(p)} [g, \Phi]  &:= \int \dd^\dimM x \sqrt{-g}\  \mathcal{O}^{(p)} (g,\Phi),
\end{align}
%--------------------------------------------------------
where each of the terms $\mathcal{O}^{(p)}(g,\Phi)$ contain $p$ powers of the matter fields
%--------------------------------------------------------
\begin{align}
    \mathcal{O}^{(p)} (g, \lambda \Phi) = \lambda^p  \mathcal{O}^{(p)} (g,  \Phi).
\end{align}
%--------------------------------------------------------
The index $N$ could in principle extend up to infinity, as it does generically in Effective Field Theories (EFT). Thus, our results will be valid, at least, for every theory that admits a Lorentz-Invariant EFT expansion. Let us perform a perturbative expansion of this action:
%--------------------------------------------------------
\begin{align}
    S[\bar{g}+\lambda h, \bar{\Phi} + \lambda \phi ] = \sum_{n = 0}^{\infty} \lambda^n \left( S_{\text{g}}^{(n)} [\bar{g},h] + S_{\text{M}}^{(n)} [\bar{g},h,\bar{\Phi},\phi] \right).
\end{align}
%--------------------------------------------------------
The gravitational parts of the action $S_{\text{g}}^{(n)}$ has the same structure as Eq.~\eqref{Eq:Partial_Actions} in Section~\ref{Subsec:Metric_Vacuum}. The matter partial actions, on the other hand, can be expressed as
%------------------------------------------------
\begin{align}
    S_{\text{M}}^{(n)} [\bar{g}, h, \bar{\Phi}, \phi]= \frac{1}{n!} \frac{\dd^n}{\dd \lambda^n} S_{\text{M}} [\bar{g} + \lambda h, \bar{\Phi}+\lambda \phi ] \bigg\rvert_{\lambda = 0}.
\end{align}
%------------------------------------------------
If we fix a vanishing matter background, the result can be expressed in terms of $\mathcal{A}_{\text{M}}^{(p)}$,
%--------------------------------------------------------
\begin{align}
    S_{\text{M}}^{(n)} [\bar{g}, h, \bar{\Phi} = 0,\phi]=\frac{1}{n!} \frac{\dd^n}{\dd \lambda^n}  \sum_{p=0}^N \lambda^p  \mathcal{A}_{\text{M}}^{(p)} [\bar{g} + \lambda h,  \phi] \bigg\rvert_{\lambda = 0},
\end{align}
%--------------------------------------------------------
where we have used $p$-th degree homogeneity of $\mathcal{O}^{(p)}$ when rescaling the matter fields with a constant. By applying explicitly the derivatives we find that terms with $p>n$ vanish, whereas terms with $p<n$ contribute nontrivially, yielding
%--------------------------------------------------------
\begin{align}
    \frac{1}{n!} \frac{\dd^n}{\dd \lambda^n}  \sum_{p=0}^N \lambda^p  \mathcal{A}_{\text{M}}^{(p)} [\bar{g} + \lambda h,  \phi] \bigg\rvert_{\lambda = 0} = \sum_{p=0}^n \frac{1}{(n-p)!} \frac{\dd^{n-p}}{\dd \lambda^{n-p}}   \mathcal{A}_{\text{M}}^{(p)} [\bar{g} + \lambda h,  \phi] \bigg\rvert_{\lambda = 0},
\end{align}
%--------------------------------------------------------
which, following the same logic as in Section \ref{Subsec:Metric_Vacuum},  can be rewritten in terms of variations with respect to the background metric as
%--------------------------------------------------------
\begin{align}
    & \sum_{p=0}^n \frac{1}{(n-p)!} \frac{\dd^{n-p}}{\dd \lambda^{n-p}}   \mathcal{A}_{\text{M}}^{(p)} [\bar{g} + \lambda h,  \phi] \bigg\rvert_{\lambda = 0} \nonumber \\
    & \qquad\qquad =\sum_{p=0}^n \frac{1}{(n-p)!} \left[ \int \dd^\dimM x\  h^{\mu\nu}\frac{\delta }{\delta \bar{g}^{\mu\nu} (x) } \right]^{n-p} \mathcal{A}_{\text{M}}^{(p)} [\bar{g} + \lambda h,  \phi] \bigg\rvert_{\lambda = 0},
\end{align}
%--------------------------------------------------------
so that we can write the matter action around $\bar{\Phi}=0$ as
%------------------------------------------------
\begin{align}
    S_{\text{M}}[\bar{g} + \lambda h, \lambda \phi ] = \sum_{n=0}^{\infty} \lambda^n \left[\sum_{p=0}^n \frac{1}{(n-p)!} \left( \int \dd^\dimM x\ h^{\mu\nu}\frac{\delta }{\delta \bar{g}^{\mu\nu} (x) } \right)^{n-p} \mathcal{A}_{\text{M}}^{(p)} [\bar{g} + \lambda h,  \phi] \right]_{\lambda = 0}.
\end{align}

%------------------------------------------------
The $n$-th order piece of energy-momentum tensor, which contains $n$ powers of $\lambda$, can be rewritten as a sum of a term coming from the gravitational sector and another term coming from the matter sector:
%------------------------------------------------
\begin{align}
    t^{(n)}_{\mu \nu} = t^{(n)}_{\text{g}\, \mu \nu} + t^{(n)}_{\text{M}\, \mu \nu},
\end{align}
%------------------------------------------------
where the piece corresponding to the gravitational energy-momentum tensor $t^{(n)}_{\text{g}\,\mu \nu}$ is given by Eq.~\eqref{Eq:Gravitational_EMTensor}, and the matter energy-momentum tensor is given by
%------------------------------------------------
\begin{equation}
     t^{(n)}_{\text{M}\,\mu \nu} := - \frac{\lambda^{n}}{\sqrt{-\bar{g}}} \frac{\delta S^{(n)}_{\text{M}} [\bar{g}, h, \bar{\Phi}, \phi]}{\delta \bar{g}^{\mu\nu}}. \label{eq:tMnvanishingmatter}\\
\end{equation}
%------------------------------------------------
In particular, around vanishing matter background, the latter can be computed from $\mathcal{A}_{\text{M}}^{(p)}$ as follows: 
%------------------------------------------------
\begin{align}
   t^{(n)}_{\text{M}\,\mu \nu}|_{\bar{\Phi}=0}
   &=- \frac{\lambda^{n}}{\sqrt{-\bar{g}}}\frac{\delta}{\delta \bar{g}^{\mu\nu}}\left[\sum_{p=0}^n \frac{1}{(n-p)!} \left( \int \dd^\dimM x\ h^{\mu\nu}\frac{\delta }{\delta \bar{g}^{\mu\nu} (x) } \right)^{n-p} \mathcal{A}_{\text{M}}^{(p)} [\bar{g} + \lambda h,  \phi] \right]_{\lambda = 0}
\end{align}
%------------------------------------------------
Notice that now every order of $\lambda$ mixes terms $\mathcal{O}^{(p)}$ and $\mathcal{O}^{(q)}$ with $p+q=n$, so that the $n$-th order matter energy-momentum tensor receives contributions from terms in the Lagrangian containing $n$ fields, but also contributions that mix terms with $p$ and $q$ fields provided $p+q=n$.

~

%------------------------------------------------
\paragraph*{\textbf{Example: Scalar field.}}
%------------------------------------------------

Let us consider a simple example for illustrative purposes: the minimally coupled $\phi^4$ theory for the matter sector and the Einstein-Hilbert term for the gravitational piece of the action. The matter sector of the theory is described by
%------------------------------------------------
\begin{align}
    S_{\text{M}}[g, \Phi] = -  \int \dd^\dimM x \sqrt{-g} \left[ \frac{1}{2} (g^{\mu\nu} \partial_\mu \Phi \partial_\nu \Phi + m^2 \Phi^2) + \frac{\beta}{4!} \Phi^4\right],
    \label{Eq:gphi4action}
\end{align}
%------------------------------------------------
and can be expressed simply in terms of two operators $\mathcal{O}^{(p)} (g, \Phi) $
%------------------------------------------------
\begin{align}
    & \mathcal{O}^{(2)} = - \frac{1}{2} g^{\mu \nu} \partial_{\mu} \Phi \partial_{\nu} \Phi - \frac{1}{2} m^2 \Phi^2, \\
    & \mathcal{O}^{(4)} = - \frac{\beta}{4!} \Phi^4.
\end{align}
%------------------------------------------------
We can now simply expand on top of a background solving the gravitational vacuum equations, so that the matter background is given by vanishing $\Phi$. Due to the vanishing background, terms containing $n$-powers of the fields will enter only at order $n$ in the stress energy tensor. Hence, although both the $m^2 \Phi^2$ and the $\beta \Phi^4$ terms can be regarded just as the potential term, they are not equal from the point of view of the bootstrapping. This is a consequence of being consistent in matching the $n$-th order gravitational perturbations with the $n$-th order matter perturbations, since they are not independent expansions, as can be seen by looking at the orders in our bookkeeping parameter $\lambda$. Indeed, for the matter action in Eq.~\eqref{Eq:gphi4action} one first has to take into account the identity
%------------------------------------------------
\begin{align}
    \sqrt{-g}&=\sqrt{-\bar{g}} \bigg[ 1 -\frac{1}{2}\lambda[\mathbf{h}] + \frac{1}{8}\lambda^2 (2[\mathbf{h}^2]+[\mathbf{h}]^2) - \frac{1}{48}\lambda^3\left(8[\mathbf{h}^3]+6[\mathbf{h}][\mathbf{h}^2]+[\mathbf{h}]^3 \right)\nonumber\\
    &\qquad\qquad + \frac{1}{384}\lambda^4\left(48[\mathbf{h}^4] + 32[\mathbf{h}][\mathbf{h}^3] +12[\mathbf{h}^2]^2 +12[\mathbf{h}]^2[\mathbf{h}^2] +[\mathbf{h}]^4\right)+\mathrm{O}(\lambda^5)\bigg],
\end{align}
%------------------------------------------------
where we have introduced the matrix $\mathbf{h}$ with components $(\mathbf{h}^\mu{}_\nu) := h^{\mu\rho}\bar{g}_{\rho\nu}$ and the symbol $[\mathbf{X}]$ as an abbreviation for the trace of a certain matrix $\mathbf{X}$. With this in mind, it is straightforward to derive:
%------------------------------------------------
\begin{align}
    S^{(0)}_\text{M}[\bar{g},h,\bar{\Phi} = 0,\phi]&=S^{(1)}_\text{M}[\bar{g},h,\phi]=0\,,\\
    S^{(2)}_\text{M}[\bar{g},h,\bar{\Phi} = 0,\phi]&=-\frac{1}{2}\int \dd^\dimM x\ \sqrt{-\bar{g}}\ (\bar{g}^{\mu\nu}\partial_\mu\phi \partial_\nu\phi+m^2\phi^2)\,,\\
    S^{(3)}_\text{M}[\bar{g},h,\bar{\Phi} = 0,\phi]&=\quad \frac{1}{4}\int \dd^\dimM x\ \sqrt{-\bar{g}}\ \Big[[\mathbf{h}](\bar{g}^{\mu\nu}\partial_\mu\phi \partial_\nu\phi+m^2\phi^2) - 2h^{\mu\nu}\partial_\mu\phi\partial_\nu\phi\Big]\,,\\
    S^{(4)}_\text{M}[\bar{g},h,\bar{\Phi} = 0,\phi]&=-\frac{1}{16}\int \dd^\dimM x\ \sqrt{-\bar{g}}\ \Big[([\mathbf{h}]^2+2[\mathbf{h}^2])(\bar{g}^{\mu\nu}\partial_\mu\phi \partial_\nu\phi+m^2\phi^2) \nonumber\\
    &\qquad\qquad\qquad\qquad\qquad+ \frac{\beta}{3}\phi^4 -4[\mathbf{h}]h^{\mu\nu}\partial_\mu\phi\partial_\nu\phi\Big] \,,
\end{align}
%------------------------------------------------
Where the contributions from $\mathcal{O}^{(p)}$ and $\mathcal{O}^{(q)}$ terms such that $p+q'=n$ can be seen explicitly at each of the computed order.
%------------------------------------------------
\subsection{Gravity with torsion}
\label{Subsec:Metric_Torsion}
%------------------------------------------------

Let us move on to include torsion into the picture. We will show that diffeomorphism-invariant theories of gravity in which both the metric and the torsion tensor are dynamical variables, bootstrap in a similar way. As we discussed in the introduction, we expect the source of the torsion perturbations to be the spin-density current, obtained as the variation of the action with respect to the torsion tensor. The discussion in this section will be similar to the one above since, to some extent, torsion behaves as some additional matter fields with possible non-minimal couplings. Consider a generic action that we can again break in two terms, one containing only the metric and one containing the metric, the torsion tensor (through the contorsion tensor $K_{\mu\nu}{}^\rho$), and possibly matter fields
%--------------------------------------------------------
\begin{align}
    S[g,K, \Phi ] = S_{\text{g}} [g] + S_{\text{F}} [g, K, \Phi ].
\end{align}
%--------------------------------------------------------
Following the same logic that we have followed in previous sections, we will expand this action in terms of a background solution:
%------------------------------------------------
\begin{align}
     g^{\mu\nu} &= \bar{g}^{\mu\nu} + \lambda h^{\mu\nu}, \\
     K_{\mu\nu}{}^\rho &= \bar{K}_{\mu\nu}{}^\rho + \lambda k_{\mu\nu}{}^\rho ,\\
     \Phi^A &= \bar{\Phi}^{A} + \lambda \phi^A, 
\end{align}
%------------------------------------------------
where in principle we can choose any solution to the equations of motions, {\it i.e.} any $\{\bar{g}^{\mu\nu},\,\bar{K}_{\mu\nu}{}^\rho,\,\bar{\Phi}^{A}\}$ such that
%--------------------------------------------------------
\begin{align}
    & \frac{\delta S[g, K, \Phi ]}{\delta g^{\mu\nu}} \bigg\rvert_{g^{\mu\nu} = \bar{g}^{\mu\nu}} = 0, \\
    & \frac{\delta S[g, K, \Phi ]}{\delta K_{\mu\nu}{}^\rho} \bigg\rvert_{ K_{\mu\nu}{}^\rho =  \bar{K}_{\mu\nu}{}^\rho }
    = 0,\\
    & \frac{\delta S[g, K, \Phi ]}{\delta \Phi^A} \bigg\rvert_{ \Phi^A = \bar{\Phi}^A} = 0.
\end{align}
%--------------------------------------------------------
If we again perform a decomposition
%--------------------------------------------------------
\begin{equation}
    S_{\text{F}}[\bar{g} + \lambda h, \bar{K}+\lambda k, \bar{\Phi}+ \lambda \phi] = \sum_{n=0}^{\infty} \lambda^n S^{(n)}_{\text{F}}[\bar{g}, h, \bar{K}, k, \bar{\Phi}, \phi],
\end{equation}
%--------------------------------------------------------
the general recursive formula \eqref{Eq:Key_Bootstrap_Gen} shows that the source of the equations of motion for $n$-th order torsion perturbations is the spin-density current of $(n-1)$-th order:
%------------------------------------------------
\begin{equation}
    \frac{\delta S^{(n)}_{\text{F}}[\bar{g}, h, \bar{K} ,k, \bar{\Phi},\phi]}{\delta k_{\mu\nu}{}^{\rho}} =\frac{\delta S^{(n-1)}_{\text{F}}[\bar{g}, h, \bar{K} ,k, \bar{\Phi},\phi]}{\delta \bar{K}_{\mu\nu}{}^{\rho}}.\label{eq:recursiveSkK}
\end{equation}
%------------------------------------------------
Indeed, if we define
%------------------------------------------------
\begin{equation}
    s^{(n)}_{\text{F}}{}^{\mu\nu}{}_{\rho} :=  \frac{\lambda^{n}}{\sqrt{-\bar{g}}} \frac{\delta S^{(n)}_{\text{F}} [\bar{g}, h, \bar{K}, k, \bar{\Phi}, \Phi ]}{\delta \bar{K}_{\mu\nu}{}^{\rho}},
\end{equation}
%------------------------------------------------
then Eq.~\eqref{eq:recursiveSkK} can be rewritten
%------------------------------------------------
\begin{equation}
    \frac{\lambda^n}{\sqrt{-\bar{g}}}\frac{\delta S^{(n)}_{\text{F}}[\bar{g}, h, \bar{K} ,k, \phi]}{\delta k_{\mu\nu}{}^{\rho}} = \lambda s^{(n)}_{\text{F}}{}^{\mu\nu}{}_{\rho},
\end{equation}
%--------------------------------------------------------
in complete analogy with~\eqref{Eq:Bootstrap_Eq_EMT}.

Let us particularize now to a trivial background for the matter fields: $\bar{\Phi}^A = 0$. Similarly as in the previous section, the metric and the contorsion backgrounds cannot be evaluated in the trivial ones ($\bar{g}^{\mu\nu}=\eta^{\mu\nu}$ and $\bar{K}_{\mu\nu}{}^{\rho}=0$), since the computation of the energy-momentum tensor and the spin-density current involve variations with respect to $\bar{g}^{\mu\nu}$ and $\bar{K}_{\mu\nu}{}^{\rho}$, and this requires knowledge about \textit{arbitrary} configurations close to the trivial backgrounds. We will, as before, assume that the action $S_{\text{F}}$ for matter and torsion can be split as follows,
%--------------------------------------------------------
\begin{align}
    & S_{\text{F}} [g, K, \Phi ] = \sum_{p=2}^{N}  \mathcal{A}_{\text{F}} ^{(p)} [g, K, \Phi ], \\
    & \mathcal{A}_{\text{F}} ^{(p)} [g, K, \Phi ]  = \int \dd^\dimM x \sqrt{-g}\  \mathcal{O}^{(p)} (g,K, \Phi ),
\end{align}
%--------------------------------------------------------
where, each of the terms $\mathcal{O}^{(p)}(g,K, \Phi )$ is homogeneous of degree $p$ in the matter fields,
%--------------------------------------------------------
\begin{align}
    \mathcal{O}^{(p)} (g, K,  \lambda \Phi) = \lambda^p  \mathcal{O}^{(p)} (g,  K, \Phi).
\end{align}
%--------------------------------------------------------
Similarly as we did in the previous section, around vanishing matter background, the partial actions in the decomposition
%------------------------------------------------
\begin{equation}
    S_{\text{F}}[\bar{g} + \lambda h, \bar{K}+\lambda k, \lambda \phi] = \sum_{n=0}^{\infty} \lambda^n S^{(n)}_{\text{F}}[\bar{g}, h, \bar{K}, k, \bar{\Phi}=0,\phi] 
\label{eq:SvanishingPhi}
\end{equation}
%------------------------------------------------
are given by
%------------------------------------------------
\begin{align}
    &S^{(n)}_{\text{F}}[\bar{g}, h, \bar{K}, k, \bar{\Phi}=0,\phi]  =  \nonumber\\
    &\qquad\sum_{p=0}^n \frac{1}{(n-p)!}\left[\int \dd^\dimM x\ \left(h^{\mu\nu}\frac{\delta }{\delta \bar{g}^{\mu\nu} (x)} + k_{\mu\nu}{}^{\rho}\frac{\delta }{\delta \bar{K}_{\mu\nu}{}^{\rho} (x) }\right) \right]^{n-p} \mathcal{A}_{\text{F}}^{(p)} [\bar{g} + \lambda h, \bar{K}+\lambda k, \phi] \bigg\rvert_{\lambda = 0}.
\label{eq:SFnvanishingPhi}
\end{align}
%------------------------------------------------

%------------------------------------------------
\section{Bootstrapping theories with a vielbein}
\label{Sec:Bootstrapping_Vielbein}
%------------------------------------------------
For the bosonic matter fields explored in the previous section we necessarily require non-minimal couplings in order for the fields to be coupled to torsion. For fermionic matter, the minimal coupling prescription gives rise to a linear coupling between the axial torsion vector and a fermionic bilinear~\cite{Hehl1976, Shapiro2001}. However, Dirac spinors do not couple to the metric directly, but require the introduction of a frame field (vielbein). Hence, for the purpose of illustrating the bootstrapping procedure for the simplest kind of matter theories that couple to torsion, we need to understand the bootstrapping through the vielbein formalism.

%------------------------------------------------
\subsection{Metric theories written in terms of vielbein}
\label{Subsec:Vielbein_Metric}
%------------------------------------------------

We will begin by reproducing previous known results by bootstrapping a metric theory of gravity $S[g,\Phi]$ described in terms of the vielbein fields instead of the metric tensor. Hence, we deal with an action of the form
%------------------------------------------------
\begin{align}
     W[e, \Phi] := S[g(e),\Phi],
\end{align}
%------------------------------------------------
where we are emphasizing that the action $W$ depends on the vielbein only through the metric. The set $\Phi$ collectively denotes another set of possible fields: matter, torsion, etc. Let us expand for perturbations on top of a background in the form
%------------------------------------------------
\begin{align}
    & e^{\mu}{}_{a} = \bar{e}^{\mu}{}_{a} + \Tilde{\lambda} \epsilon^{\mu}{}_{a}, \\
    & \Phi^A = \bar{\Phi}^A + \Tilde{\lambda} \phi^A,
\end{align}
%------------------------------------------------
which translates into an expansion of the action as
%------------------------------------------------
\begin{align}
    W[e, \Phi] = \sum_{n=0}^{\infty} \tilde{\lambda} ^n W^{(n)} [\bar{e}, \epsilon, \bar{\Phi},\phi].
\end{align}
%------------------------------------------------

From our analysis in Appendix \ref{App:General_Fields_funidentity}, it is clear that we can find the recursive relation
%------------------------------------------------
\begin{align}
    \frac{\delta W^{(n)} [\bar{e}, \epsilon, \bar{\Phi},\phi] }{\delta \epsilon^{\mu}{}_{a}}  = \frac{\delta W^{(n-1)} [\bar{e}, \epsilon, \bar{\Phi},\phi] }{\delta \bar{e}^{\mu}{}_{a} },
\end{align}
%------------------------------------------------
where the right hand side is understood as the energy-momentum tensor at order $n-1$ in the vielbein formulation. To be more precise, we introduce the following notation for such partial energy-momentum tensors
%------------------------------------------------
\begin{equation}
    \mathfrak{t}^{(n)}{}_{\mu }{}^a := \frac{\tilde{\lambda}^n}{\abs{\bar{e}}}\frac{\delta W^{(n)} [\bar{e}, \epsilon, \bar{\Phi},\phi] }{\delta \bar{e}^{\mu}{}_{a} },
\end{equation}
%------------------------------------------------
in complete analogy with Eq.~\eqref{Eq:Gravitational_EMTensor}.

We thus see that for every theory in which the gravitational field is described by a vielbein, gravitational perturbations couple to their own energy-momentum tensor, as we have not used that the vielbein enters only through the metric up to this point. Note, however, that in the case where the vielbein enters through a metric tensor, we have a problem to match the $\mathfrak{t}^{(n)}{}_{\mu}{}^{a}$ with the $t^{(n)}_{\mu \nu}$ from the previous section. The main reason for this is that the two expansions are different (notice that we have emphasized this using a different parameter $\tilde{\lambda}$) and vielbein perturbations cannot be trivially matched to metric perturbations order by order in the expansion. Indeed, if we construct the sum of partial actions up to a given accuracy
%-------------------------------------------------------
\begin{align}
    W = \sum_{n=0}^N \tilde{\lambda}^n W^{(n)} + \order{\tilde{\lambda}^{N+1}},
\end{align}
%-------------------------------------------------------
the associated energy-momentum tensor in the vielbein formulation would be the sum of all partial contributions:
%-------------------------------------------------------
\begin{align}
    \sum_{n=0}^N \mathfrak{t}^{(n)}{}_{\mu }{}^a + \order{\tilde{\lambda}^{N+1}}= \frac{1}{\abs{\bar{e}}}\frac{\delta  }{\delta \bar{e}^{\mu}{}_{a} } \sum_{n=0}^N \tilde{\lambda}^n W^{(n)} + \order{\tilde{\lambda}^{N+1}}.
\end{align}
%-------------------------------------------------------
To compare with the partial actions $S^{(n)}$, we would need to multiply by the inverse vielbein $e_\nu{}^{a}$ which, from the point of view of the metric expansion, is also a series in $\lambda$. Making the identification between the two series would require to  make the expansion of vielbein in terms of the $h^{\mu \nu}$ field to match the tensors $\mathfrak{t}$ and $t$, which does not appear to be possible in a systematic manner. In other words, it is not straightforward to connect the \textit{partial} energy-momentum tensors computed in the metric formulation versus those obtained in the vielbein formulation. However, at the full non-linear order, both approaches should agree, as a direct consequence of the identity \eqref{Eq:Key_Bootstrap_Gen}, which we derive in Appendix~\ref{App:General_Fields_funidentity}. To illustrate explicitly this point, consider a scalar of the form
%-------------------------------------------------------
\begin{align}
    \mathcal{S} = \mathcal{M}_{\mu \nu} g^{\mu \nu} = \mathcal{M}_{\mu \nu} \bar{g}^{\mu \nu} + \lambda \mathcal{M}_{\mu \nu} h^{\mu \nu},
\end{align}
%-------------------------------------------------------
where $\mathcal{M}_{\mu \nu}$ is assumed to be independent of the metric. If we express it in terms of the vielbein, we clearly see that the series is different since it ends up at one order more
%-------------------------------------------------------
\begin{align}
    \mathcal{S} = \mathcal{M}_{\mu \nu} e^{\mu}{}_{a} e^{\nu}{}_{b}\eta^{a b} = \mathcal{M}_{\mu \nu} \bar{e}^{\mu}{}_{a} \bar{e}^{\nu}{}_{b}\eta^{a b} + 2 \tilde{\lambda} \mathcal{M}_{\mu \nu} \bar{e}^{\mu}{}_{a} \epsilon^{\nu}{}_{b}\eta^{a b} + \tilde{\lambda}^2 \mathcal{M}_{\mu \nu} \epsilon^{\mu}{}_{a} \epsilon^{\nu}{}_{b}\eta^{a b}\,.
\end{align}
%-------------------------------------------------------
This example allows us to illustrate the point that the two expansions are different: whereas the one in terms of the metric contains only up to $\lambda$ terms, the one in terms of the vielbein contains terms up to $\tilde{\lambda}^2$. However, although we could not find a closed expression matching the two expansions and the energy-momentum tensors at different orders, we have shown that, within the vielbein formalism, vielbein perturbations bootstrap by coupling to its own energy-momentum tensor.

%------------------------------------------------
\subsection{Matter coupled to torsion}
\label{Subsec:Vielbein_Torsion}
%------------------------------------------------

At this point, extending the arguments in the previous section to the case of considering matter and torsion in the vielbein formalism is straightforward. We will encode the torsion again into the contorsion tensor with two Latin indices, $K_{ \mu a b}$, noting that the last index can be kept lowered, since the Minkowski metric is not affected by functional variations. Let us consider again an action of the form
%------------------------------------------------
\begin{align}
    S[e,K, \Phi ] = S_{\text{g}} [g(e)] + S_{\text{F}} [e, K, \Phi ].
\end{align}
%------------------------------------------------
Once more, we will assume that the matter action can be written as
%--------------------------------------------------------
\begin{align}
    & S_{\text{F}} [e, K, \Phi ] = \sum_{p=2}^{N}  \mathcal{A}_{\text{F}} ^{(p)} [e, K, \Phi ], \\
    & \mathcal{A}_{\text{F}} ^{(p)}  = \int \dd^\dimM x \sqrt{-g}\  \mathcal{O}^{(p)} (e,K, \Phi ),
\end{align}
%--------------------------------------------------------
where $N$ can extend up to infinity and each of the terms $\mathcal{O}^{(p)}(e,K, \Phi )$ is homogeneous of degree $p$, satisfying 
%---------------------------------
\begin{equation}
    \mathcal{O}^{(p)} (e, K, \tilde{\lambda} \Phi) = \tilde{\lambda}^p  \mathcal{O}^{(p)} (e,  K, \Phi ).
\end{equation}
%--------------------------------------------------------
We expand again on top of a background that solves the equations of motion and which is trivial for the matter. Following the above arguments, we can expand the action around a vanishing matter background as (compare with Eqs. \eqref{eq:SvanishingPhi}-\eqref{eq:SFnvanishingPhi})
%------------------------------------------------
\begin{equation}
    S_{\text{F}}[\bar{e} + \tilde{\lambda}\epsilon, \bar{K}+\tilde{\lambda} k, \tilde{\lambda} \phi] = \sum_{n=0}^{\infty} \lambda^n S^{(n)}_{\text{F}}[\bar{e} , \epsilon, \bar{K}, k, \bar{\Phi}=0, \phi]    
\end{equation}
%------------------------------------------------
with
%------------------------------------------------
\begin{align}
    &S^{(n)}_{\text{F}}[\bar{e}, \epsilon, \bar{K}, k, \bar{\Phi}=0,\phi]  =  \nonumber\\
    &\qquad\sum_{p=0}^n \frac{1}{(n-p)!}\left[\int \dd^\dimM x\ \left(\epsilon^{\mu}{}_{a} \frac{\delta }{\delta \bar{e}^{\mu}{}_{a} (x)} + k_{\mu ab}\frac{\delta }{\delta \bar{K}_{\mu ab} (x) }\right) \right]^{n-p} \mathcal{A}_{\text{F}}^{(p)} [\bar{e} + \tilde{\lambda}\epsilon, \bar{K}+\tilde{\lambda} k, \phi] \bigg\rvert_{\tilde{\lambda} = 0}.
\end{align}
%------------------------------------------------
Again, the $n$-th order energy-momentum tensor contains two pieces,
%------------------------------------------------
\begin{align}
    \mathfrak{t}^{(n)}_{\text{}}{}_{\mu}{}^a = \mathfrak{t}^{(n)}_{\text{g}}{}_{\mu}{}^a + \mathfrak{t}^{(n)}_{\text{F}}{}_{\mu}{}^a, 
\end{align}
%------------------------------------------------
where the piece corresponding to the gravitational energy-momentum tensor $\mathfrak{t}^{(n)}_{\text{g}}{}_{\mu}{}^a$ is the one from the previous section, and the remaining contribution is
%------------------------------------------------
\begin{align}
    \mathfrak{t}^{(n)}_\text{F}{}_{\mu}{}^a = - \frac{\tilde{\lambda}^n}{\abs{\bar{e}}} \frac{\delta S^{(n)}_{\text{F}} [\bar{e}, \epsilon, \bar{K}, k, \bar{\Phi}, \phi]}{\delta \bar{e}^{\mu}{}_{a} }.
\end{align}
%------------------------------------------------
As before, the  $\tilde{\lambda}$ expansion is such that terms $\mathcal{O}^{(p)}$ and $\mathcal{O}^{(q)}$, with $p+q=n$, get mixed at order $n$.

Analogously, one can introduce the $n$-th order spin-density current, which has only the contribution coming from $S_{\text{F}}$ and is given by:
%------------------------------------------------
\begin{equation}
    \mathfrak{s}^{(n)}{}^{\mu ab} = \mathfrak{s}^{(n)}_{\text{F}}{}^{\mu ab} := \frac{\tilde{\lambda}^n}{|\bar{e}|}\frac{\delta S^{(n)}_{\text{F}}[\bar{e},\epsilon, \bar{K},k, \bar{\Phi}, \phi]}{\delta \bar{K}_{\mu ab}}.
\end{equation}
%------------------------------------------------

\paragraph*{\textbf{Example: Dirac action with torsionful derivative.}}
%------------------------------------------------
Let us consider the example of a Dirac fermion $\Psi$ minimally coupled to a Poincaré gauge gravity theory ({\it i.e.}, the connection is metric-compatible) in four dimensions:
%-------------------------------------------------------
\begin{align}
    S_{\textrm{Dirac}}[e,K,\Psi,\CPsi] = \int \dd^4 x \abs{e} \left[ \frac{\ii}{2} e^\mu{}_c \left(\CPsi \gamma^c \GCD_\mu \Psi- \overline{\GCD_\mu\Psi} \gamma^c \Psi \right)  - m \CPsi \Psi  \right],\label{Eq:DiracActionTorsion}
\end{align}
%-------------------------------------------------------
where note that we are representing the Dirac conjugate with a longer bar, as $\CPsi$, and it should not be confused with the background value of the field, $\bar{\Psi}$. In our signature convention, the covariant derivative acts on spinors as
%-------------------------------------------------------
\begin{align}
   \GCD_\mu\Psi
   &:= \partial_{\mu}\Psi + \frac{1}{4} \mathring{\omega}_{\mu a b} \gamma^{[a} \gamma^{b]}\Psi + \frac{1}{4} K_{\mu a b} \gamma^{[a} \gamma^{b]} \Psi\nonumber \\
   &=  \mathring{\nabla}_\mu\Psi + \frac{1}{4} K_{\mu a b} \gamma^{[a} \gamma^{b]}\Psi .
\end{align}
%-------------------------------------------------------
The above Dirac action can then be rewritten as
%-------------------------------------------------------
\begin{align}
    S_{\textrm{Dirac}} &= \int \dd^4 x \abs{e} \left[ \frac{\ii}{2} e^\mu{}_c\left(\CPsi\gamma^c\mathring{\nabla}_\mu \Psi- \overline{\mathring{\nabla}_\mu \Psi} \gamma^c \Psi \right)  - m \CPsi \Psi -\frac{1}{4}  K_{\mu a b}e^\mu{}_c\,(\ii \CPsi \gamma^{[a}\gamma^{b}\gamma^{c]}\Psi) \right]\,,
\end{align}
%-------------------------------------------------------
where we have used $\gamma^c \gamma^{[a}\gamma^{b]}+ \gamma^{[a}\gamma^{b]}\gamma^c = 2 \gamma^{[a}\gamma^b\gamma^{c]}$. Here we can clearly see how the contorsion couples to the (Hodge dual of) the axial current:
%-------------------------------------------------------
\begin{equation}
    \ii\CPsi\gamma^{[a}\gamma^{b}\gamma^{c]}\Psi=\varepsilon^{dabc} \CPsi\gamma_d\gamma_5\Psi,
\end{equation}
%-------------------------------------------------------
where $\gamma_5:=\ii \gamma^0\gamma^1\gamma^2\gamma^3$ and $\varepsilon^{dabc}$ is the Levi-Civita tensor of the flat metric. Indeed, only its totally antisymmetric part $K_{[c a b]} := K_{\mu [a b}e^\mu{}_{c]}$ couples to the spin current. Since the field $K_{\mu a b}$ does not enter with derivatives, the bootstrapping closes after the first iteration. Indeed, the total spin-density around the background with vanishing spinor is given by its lowest order,
%-------------------------------------------------------
\begin{equation}
    \sum_{n=0}^\infty \mathfrak{s}^{(n)}{}^{\mu ab}|_{\bar{\Psi}=0} = \mathfrak{s}^{(2)}{}^{\mu ab}|_{\bar{\Psi}=0}
    =  \frac{1}{|\bar{e}|} \frac{\delta S_{\textrm{Dirac}}[\bar{e},\bar{K},\psi,\overline{\psi}]}{\delta\bar{K}_{\mu a b}} = \frac{\ii}{4} \bar{e}^{\mu}{}_{c} \overline{\psi}\gamma^{[c}\gamma^{a}\gamma^{b]}\psi\,,
\end{equation}
%-------------------------------------------------------
where $\psi$ is the deviation with respect to the trivial spinor background. This example illustrates that the presence of torsion does not spoil the bootstrapping of the vielbein perturbations that couple to the matter and gravitational energy-momentum tensor. Furthermore, as we expected, the torsion field is coupled to the spin-density current. All of this occurs in a self-consistent way, order by order in the expansion parameter $\tilde{\lambda}$.
%-------------------------------------------------------
\section{The bootstrapping of nonmetricity}
\label{Sec:Nonmetricity}
%-------------------------------------------------------

In this section we will discuss how the conclusions of the previous sections can be extended to the case in which the matter fields are coupled to the nonmetricity. Nevertheless, there is a subtle point when nonmetricity is introduced into the picture regarding the interpretation of the source for the equations of motion. Usually, the field theories that we take as starting point are Poincaré invariant theories. This implies the conservation of the canonical energy-momentum and angular momentum currents, the latter involving a promotion of the type $\partial\to \hat{\nabla}$, where $\hat{\nabla}$ is a Lorentz connection (metric-compatible).

In case of having non-trivial nonmetricity, the corresponding connection $\omega_{\mu a}{}^b$ will be a ${\rm GL}(\dimM,\mathbb{R})$-connection instead of a Lorentz one. This leads to difficulties in interpreting the right-hand side of the equation of motion for the nonmetricity as a canonical current like the energy-momentum tensor for the metric equations and the spin-density current for the torsion equations. In other words, the source of the equations of motion obtained by varying 
the action with respect to the nonmetricity order by order, which is called the dilation-shear current~\cite{Hehl1995}, cannot be alternatively computed as a Noether current in the canonical approach. It can only be computed through Hilbert's prescription. In order to be able to obtain such a current from the canonical approach, we would need the theory to be invariant under the whole ${\rm GL}(\dimM,\mathbb{R})$, and the general theories that we consider are only Poincar\'e invariant as we have already advanced. Otherwise, there are no other conserved currents besides the angular momentum and the energy-momentum tensor. This issue becomes specially problematic if we want to deal with spinors, since they do not have well definite transformation properties under ${\rm GL}(\dimM,\mathbb{R})$, but only under the Lorentz subgroup. Indeed, this is a consequence of ${\rm GL}(\dimM,\mathbb{R})$ not admitting finite spinor representations, but only infinite dimensional ones \cite{Neeman1978}.  

If we blindly apply the bootstrapping procedure defining the currents only in terms of the Hilbert prescription, the master equation for the perturbations still applies and the nonmetricity would couple order by order to the dilation-shear current. Everything is completely parallel to the already studied case of torsion. However, the absence of a clear interpretation for the source obscures the interpretation of the procedure.

%------------------------------------------------
\section{Bootstrapping Unimodular metric-affine gravity}
\label{Sec:Unimodular}
%------------------------------------------------

All of our discussion up to this point has been based on background-independent theories, {\it i.e.}, theories that exhibit diffeomorphism invariance. However, we can now consider theories in which one has a privileged background volume form that will enter our action, although it will not be a dynamical field with respect to which we perform variations. These theories are the higher order derivative generalizations of the so-called unimodular gravity, and are in general formulated easily in terms of their general relativistic counterparts, see~\cite{Carballo2022} for a recent review. The symmetry group here is a semi-direct product of Weyl transformations and transverse diffeomorphisms (those preserving the volume form introduced). 

Consider for instance a general relativistic action described in terms of the action $S_{\text{GR}}[g]$. To promote this to a unimodular-like theory, we need to introduce the privileged background volume form $\boldsymbol{\omega} = \omega (x)~\dd x^0 \wedge ... \wedge \dd x^{\dimM-1}$. The way to do it, is to introduce an auxiliary metric $\tilde{g}_{\mu \nu}$ defined in terms of our dynamical metric $g_{\mu \nu}$ and the volume form: 
%------------------------------------------------
\begin{equation}
    \tilde{g}_{\mu \nu} (g) = g_{\mu \nu} \left( \frac{\omega^2}{\abs{g}} \right)^{\frac{1}{\dimM}}.
    \label{Eq:Auxiliary_Metric}
\end{equation}
%------------------------------------------------
With this, we would write an action for the dynamical variable $g_{\mu \nu}$, taking advantage of this auxiliary metric as
%------------------------------------------------
\begin{equation}
    S_{\text{UG}}[g] = S_{\text{GR}} [\tilde{g}(g)].
\end{equation}
%------------------------------------------------
One can work out the equations of motion for this theory and one immediately finds out that the equations that one reaches are the traceless part of the equations of motion for the General Relativistic counterpart~\cite{Carballo2022}. Explicitly, we have the following
%------------------------------------------------
\begin{equation}
   \frac{\delta S_{\text{UG}}}{\delta g^{\mu \nu}} = E_{\mu \nu} [\tilde{g}(g)] - \frac{1}{\dimM} \tilde{g}_{\mu \nu} \tilde{g}^{\alpha \beta} E_{\alpha \beta} [\tilde{g}(g)],
\end{equation}
%------------------------------------------------
where $E_{\mu \nu} [g] := \dfrac{\delta S_{\text{GR}}[g]}{\delta g^{\mu \nu}}$. These theories do not have the full group of diffeomorphisms as a symmetry group, but instead they have the group of transverse diffeomorphisms and Weyl-scalings as gauge symmetries. Explicitly, these transformations are:
%------------------------------------------------
\begin{align}
    & \delta_{\xi} g_{\mu \nu} = 2 \CDg_{( \mu} \xi_{\nu)}, \qquad \CDg_{\mu} \xi^{\mu} = - \frac{1}{2} \xi^{\sigma} \partial_{\sigma} \log \left( \frac{\omega^2}{\abs{g}} \right), \\
    & \delta_{\phi} g_{\mu \nu} =  \frac{1}{2} \phi g_{\mu \nu},
\end{align}
%------------------------------------------------
If one includes matter or additional fields in such a way that these gauge symmetries are preserved, {\it i.e.} we couple them through the metric $\tilde{g}_{\mu \nu} (g)$. The equations that one obtains if one adds some matter content in this way $S_{\text{M}} [g, \Phi]$, are again the traceless equations of motion of the general relativistic equations of motion, {\it i.e.},
%------------------------------------------------
\begin{align}
    E_{\mu \nu} \left[\tilde{g} (g) \right] - \frac{1}{\dimM} \tilde{g}_{\mu \nu} \tilde{g}^{\alpha \beta} E_{\alpha \beta} \left[ \tilde{g}(g)\right] = T_{\mu \nu} \left[\tilde{g} (g) \right] - \frac{1}{\dimM} \tilde{g}_{\mu \nu} \tilde{g}^{\alpha \beta} T_{\alpha \beta }\left[\tilde{g} (g) \right] .
    \label{Eq:Unimodular_Matter_Eqs}
\end{align}
%------------------------------------------------
Here the energy momentum tensor is computed through Hilbert's prescription, {\it i.e.} given $S_{\text{M}} [g,\Phi]$, the energy-momentum tensor is directly computed as
%------------------------------------------------
\begin{align}
    T_{\mu \nu} [g]= \frac{-2}{\sqrt{-g}}  \frac{\delta S_{\text{M}} [g, \Phi]}{\delta g^{\mu \nu}}.
\end{align}
%------------------------------------------------
Taking the divergence on both sides in Eq.~\eqref{Eq:Unimodular_Matter_Eqs} and performing a trivial integration, one can find that a cosmological constant appears as an integration constant, and the equations can be rewritten as their general relativistic counterpart up to this subtlety regarding the cosmological constant, a well-known properties of unimodular gravity. Further details of these theories and a detailed comparative study of them with the General Relativistic counterpart can be found in~\cite{Carballo2022}.

Let us now briefly discuss the effect of using this alternative symmetry principle to build theories and their bootstrapping. If we perform the expansion of the metric again in a background plus a perturbation of the form $g^{\mu \nu} = \bar{g}^{\mu \nu} + \lambda h^{\mu \nu}$, the equations that we proved in the appendix still hold, even in the presence of a background volume form, {\it i.e.}
%------------------------------------------------
\begin{equation}
    \frac{\delta S^{(n-1)} [\omega, \bar{g}, h]}{\delta \bar{g}^{\mu \nu} } = \frac{\delta S^{(n)} [\omega, \bar{g}, h]}{\delta h^{\mu \nu}},
\label{Eq:Bootstrap_Unimodular}
\end{equation}
%------------------------------------------------
If we now take into account that metric $g_{\mu \nu}$ always enters into the action through the $\tilde{g}_{\mu \nu}$ metric, we have that
%------------------------------------------------
\begin{equation}
    \frac{ \delta S^{(n)}[\omega, \bar{g}, h]}{\delta g^{\mu \nu}} = \frac{\delta S^{(n)} [\omega, \bar{g}, h]}{\delta \bar{g}^{\mu \nu}} - \frac{1}{\dimM} \bar{g}_{\mu \nu} \bar{g}^{\alpha \beta } \frac{\delta S^{(n)} [\omega, \bar{g}, h]}{\delta \bar{g}^{\alpha \beta}}.
\end{equation}
%------------------------------------------------
Now if we identify the energy momentum tensor from the definition above, at order $n-1$ in $\lambda$ to find the following expression
%------------------------------------------------
\begin{equation}
    \frac{\delta S^{(n)} [ \omega, \bar{g}, h]}{\delta h^{\mu \nu}}  = -  \sqrt{-\bar{g}} \left( t^{(n-1)}_{\mu \nu}   - \frac{1}{\dimM} \bar{g}_{\mu \nu} \bar{g}^{\alpha \beta} t^{(n-1)}_{\alpha \beta} \right).
\end{equation}
%------------------------------------------------
This expression is simply telling us that in this case, instead of coupling $h^{\mu \nu}$ to its own energy momentum tensor, it couples to the traceless part of its own energy-momentum tensor.

Similarly will occur for the case in which we include matter, we work with the vielbein or we include torsion and nonmetricity. The unique difference with respect to the diffeomorphism-invariant actions is that now the coupling is given to the traceless part of its energy momentum tensor.

%------------------------------------------------
\section{Conclusions}
\label{Sec:Conclusions}
%------------------------------------------------

In this work we have worked out explicitly the self-coupling problem of metric-affine theories of gravity to suitable currents. For that purpose, we first generalized the self-coupling of arbitrary metric theories of gravity done by Butcher~\emph{et al.} to include also an arbitrary matter content in section~\ref{Subsec:Metric_Matter}, showing that metric perturbations always couple to the Hilbert energy-momentum tensor. Then we worked out the self-coupling for a theory with torsion, first in the metric formalism (in Subsection~\ref{Subsec:Metric_Torsion}); and then in the vielbein formalism in Section~\ref{Sec:Bootstrapping_Vielbein}. We managed to show that the metric or vielbein perturbations still couple to the energy-momentum tensor computed {\it \`a la} Hilbert, whereas the torsion perturbations couple to the spin-density current. We also discussed what happens with nonmetricity in Section~\ref{Sec:Nonmetricity} and found similar results. The coupling in this case is to the shear-dilation current. However, in general such current does not admit a definition in terms of a canonical current (generic theories are not invariant under the whole General Linear group) and hence it obscures the interpretation of the result for general theories. Finally, we explained how our analysis generalize straightforwardly if we consider a unimodular gravity version of the theories examined in the paper in Section~\ref{Sec:Unimodular}.

We have also taken the opportunity in this paper to clarify some doubts and misconcepctions that we have found in the literature. First of all, regarding the results found by Butcher \emph{et al.} in~\cite{Butcher2009}. There, they concluded that Einstein's General Relativity does not self-couple to its own energy-momentum tensor. This statement relies on the fact that if one begins with the flat-spacetime Fierz-Pauli action in the quadratic Lagrangian replacing the partial derivatives by covariant derivatives, one does not succeed when performing the bootstrapping. To succeed, one needs to add some non-minimal couplings to compute the energy-momentum tensor through Hilbert's prescription, or superpotential terms if done in the canonical way. Such terms arise naturally by expanding Einstein-Hilbert action on top of an arbitrary background but, being non-minimal, they automatically vanish when one particularizes to flat spacetime. We have argued that those terms manifest in the form of identically conserved terms in the energy-momentum tensor which, although irrelevant for computing conserved charges, are needed in order to succeed in the bootstrapping procedure. In that sense, we think that the Einstein-Hilbert action is indeed a solution to the self-coupling problem although it requires to choose adequately the ambiguities present in the definition of the energy-momentum tensor to succeed. We also note that these ambiguities are harmless when working in Palatini formalism since one simply needs to add no term to the energy-momentum tensor obtained canonically, as it was shown by Deser originally~\cite{Deser1970}, although in this approach an auxiliary field is introduced in the bootstrapping but not bootstrapped. Furthermore, we have clarified the fact that higher derivative theories of gravity do indeed also bootstrap, contrary to what was pinpointed in~\cite{Deser2017}. Actually, the drawback of the analysis in~\cite{Deser2017} is that working in Palatini formalism is not equivalent to the standard metric formalism in higher derivative generalizations. Hence, if one works in a Palatini formalism, one has to analyze also the bootstrapping of the affine connection (or equivalently torsion and nonmetricity). When one does it carefully as we have done here, one finds out that the theory still couples to some currents: the metric still couple to the energy-momentum tensor and the torsion and nonmetricity to the spin-density current and the shear-dilation current. The discussion of the role of these ambiguities in the bootstrap prescription also led us to point out a possible connection between them and Lovelock theorem, in the sense that due to the uniqueness of Lovelock actions in $\dimM$-dimensions, only very particular choices of superpotentials/non-minimal couplings will lead to these theories after the bootstrapping process is finished. Indeed, from this perspective, our results suggest that hoping to bootstrap a generic linear theory without any knowledge of the completion is hoping for a miracle, since one would have to choose the right non-minimal couplings/superpotential terms to end up with the desired result. Furthermore, without knowing them a priori, it seems highly unlikely to be able to guess them from physical considerations, since they do not affect neither physical observables of the known linear theory, nor conserved charges of the full non-linear one. Of course, there are exceptions such as the usual gauge theories that do bootstrap without having to make such guesses, but this is not generally the case.

In addition to understand some questions raised in previous works on the topic, our results are interesting also for several reasons. Regarding embedding metric-affine theories of gravity into emergent approaches, the bootstrapping procedure is a key tool~\cite{Barcelo2021,Barcelo2021b,Volovik2009}. Although in emergent approaches it is easy to analyze the degrees of freedom that can arise in linearized theories around a Fermi-point, it is hard to understand the potential non-linear completions of such theories without further insights. If one performs a bootstrapping procedure, it is possible to find the theories that give rise to the propagation of the same number of degrees of freedom at the non-linear level. Hence, our findings here suggest that it is in principle possible to embed non-linear metric-affine theories of gravity in emergent approaches. Also it would be interesting to analyze the bootstrapping of GR in its alternative \emph{trinity} formulations~\cite{BeltranJimenez2019}. These analysis will be relatively simple based on the tools that we have introduced in this paper. It is worth remarking that in the symmetric teleparallel formulation, the action is diffeomorphism-invariant without needing any boundary terms. In the standard Einstein-Hilbert action, $ \Rg (g) \sim \partial \LCg (g) + \LCg(g) \LCg(g)$, the terms containing derivatives of the connection are boundary terms and they are the ones giving rise to the non-minimal couplings when expanding on top of an arbitrary background. Such non-minimal coupling seems to be absent in the symmetric teleparallel formulation, and hence the bootstrapping seems to be more straightforward.
%, as in Deser's seminal work~\cite{Deser1970}. 
We will report on this topic elsewhere.

%-------------------------------------------------------
\acknowledgments{
The authors would like to thank Jose Beltrán Jiménez and Francisco José Maldonado Torralba for useful discussions and feedback. GGM thanks the Laboratory of Theoretical Physics at the University of Tartu for its hospitality in the preparation of this work. GGM is funded by the Spanish Government fellowship FPU20/01684. GGM acknowledges financial support from the grant CEX2021-001131-S funded by MCIN/AEI/10.13039/501100011033, to Spanish Government through the project PID2020-118159GB-C43 and further support by the European Regional Development Fund under the Dora Plus scholarship grants. AJC was also supported by the Mobilitas Pluss post-doctoral grant MOBJD1035. This research was supported jointly by the European Regional Development Fund through the Center of Excellence TK133 “The Dark Side of the Universe”, the Estonian Research Council through the grant PRG356. A.D. is also supported by the NSF grants PHY-2110273, PHY-1903799, and PHY-2206557, by the RCS program of Louisiana Boards of Regents through the grant LEQSF(2023-25)-RD-A-04, and by the Hearne Institute for Theoretical Physics
}
%-------------------------------------------------------

%-------------------------------------------------------
\appendix
%-------------------------------------------------------

%-------------------------------------------------------
\section{Decomposition of a general connection}
\label{App:DecompositionConnection} 
%-------------------------------------------------------

Consider a general affinely connected metric manifold $(M,g,\Gamma)$. We introduce the torsion and the nonmetricity tensors, respectively, as follows:
%-------------------------------------------------------
\begin{align}
   \mathcal{T}_{\mu \nu}{}^\rho &:= \Gamma_{\mu\nu}{}^\rho -  \Gamma_{\nu\mu}{}^\rho\,,\\
   \mathcal{Q}_{\mu \nu \rho} &:= -\nabla_\mu g_{\nu\rho}\,.
\end{align}
%-------------------------------------------------------
With this in mind it can be shown that the general connection can be always expressed as the Levi-Civita connection plus a tensorial deviation that we split into two contributions:
%-------------------------------------------------------
\begin{equation}
    \Gamma_{\mu\nu}{}^\rho = \LCg_{\mu\nu}{}^\rho + K_{\mu\nu}{}^\rho + L_{\mu\nu}{}^\rho\,.
\end{equation}
%-------------------------------------------------------
In this decomposition we have introduced
%-------------------------------------------------------
\begin{align}
   K_{\mu\nu}{}^\rho &:= \frac{1}{2}g^{\rho\sigma}(\mathcal{T}_{\mu \nu \sigma}+\mathcal{T}_{\sigma\mu \nu }-\mathcal{T}_{\nu \sigma\mu })\,,\label{Eq:defContorsion}\\
   L_{\mu\nu}{}^\rho &:= \frac{1}{2}g^{\rho\sigma}(\mathcal{Q}_{\mu \nu \sigma}+\mathcal{Q}_{\nu \sigma\mu }-\mathcal{Q}_{\sigma\mu \nu })\,,
\end{align}
%-------------------------------------------------------
which are called contorsion and disformation tensors, respectively. They have the symmetry properties:
%-------------------------------------------------------
\begin{equation}
    K_{\mu\nu\rho}=-K_{\mu\rho\nu}\,,\qquad L_{\mu\nu}{}^\rho=L_{\nu\mu}{}^\rho\,.
\end{equation}
%-------------------------------------------------------
When working in an arbitrary orthogonal frame (with vielbein $e^\mu{}_a$), the components of the connection become:
%-------------------------------------------------------
\begin{equation}
    \omega_{\mu a}{}^{b} :=  \Gamma_{\mu\nu}{}^\rho e^\nu{}_a e_\rho{}^b - e^\nu{}_a\partial_\mu e_\nu{}^b.
\end{equation}
%-------------------------------------------------------
The decomposition into contorsion and disformation still holds in this formalism:
%-------------------------------------------------------
\begin{equation}
    \omega_{\mu a}{}^b=\mathring{\omega}_{\mu a}{}^b+K_{\mu a}{}^b+L_{\mu a}{}^b,
    \label{Eq:generalconnection2}
\end{equation}
%-------------------------------------------------------
where $\mathring{\omega}_{\mu a}{}^b$ is the contribution of the Levi-Civita connection, and
%-------------------------------------------------------
\begin{equation}
    K_{\mu a}{}^b:= e^\nu{}_a e_\rho{}^b K_{\mu \nu}{}^\rho \,,\qquad L_{\mu a}{}^b:=e^\nu{}_a e_\rho{}^b L_{\mu \nu}{}^\rho.
\end{equation}
%-------------------------------------------------------

%-------------------------------------------------------
\section{Functional properties}
\label{App:General_Fields}
%-------------------------------------------------------

Consider a generic action functional $S[\mathrm{Q}^I]$ depending on a certain family of spacetime fields that we denote collectively as $\{\mathrm{Q}^I (x) \}$. These fields will be the metric, matter fields, the vielbein and the contorsion: $\{\mathrm{Q}^I= g^{\mu \nu}, \Phi^A, \Psi^{A}, e^{\mu}{}_{a}, K_{\mu \nu}{}^{\rho} \}$. Hence, the indices $(I,J...)$ represent a placeholder for all the fields living in the manifold and their internal $(i,j...)$, spacetime $(\mu, \nu...)$ and frame indices $(a,b...)$. Moreover, let $\{\bar{\mathrm{Q}}^I\}$ be a generic background configuration. Now we evaluate the action in a configuration that deviates from such a background, $\mathrm{Q}^I = \bar{\mathrm{Q}}^I + \lambda \mathrm{q}^I$, where $\lambda$ is a dimensionless bookkeeping expansion parameter. The total action can then be rearranged as a series of the form:
%------------------------------------------------
\begin{align}
    S[\mathrm{Q}] = \sum_{n = 0}^{\infty} \lambda^n S^{(n)} [\bar{\mathrm{Q}},  \mathrm{q}],
\end{align}
%------------------------------------------------
where the partial actions $S^{(n)}$ are given by
%------------------------------------------------
\begin{align}
    S^{(n)} [\bar{\mathrm{Q}},  \mathrm{q}] = \frac{1}{n!} \frac{\dd^n}{\dd \lambda^n} S [\bar{\mathrm{Q}} + \lambda  \mathrm{q}] \bigg\rvert_{\lambda  = 0}.
    \label{Eq:Partial_Actions_Gen}
\end{align}

%-------------------------------------------------------
\subsection{Generating formula}
\label{App:General_Fields_genformula}
%-------------------------------------------------------
The derivatives with respect to $\lambda$ in \eqref{Eq:Partial_Actions_Gen} can be replaced by the standard functional derivative operator with respect to the background fields
%------------------------------------------------
\begin{align}
    \frac{\dd}{\dd \lambda} S[\bar{\mathrm{Q}} + \lambda  \mathrm{q}] = \int \dd^\dimM x \  \mathrm{q}^I (x) \frac{\delta}{\delta \bar{\mathrm{Q}}^{I} (x) } S[\bar{\mathrm{Q}} + \lambda  \mathrm{q}].
\end{align}
%-------------------------------------------------------
Here, it is important to highlight again that there is a sum in $I$ that covers all the fields of the theory and all of their indices. Furthermore, we have dropped surface integrals that arise when integrating by parts. These terms are neglected since we are assuming the perturbations $ \mathrm{q}^I$ of all the fields to have compact support.\footnote{
    This condition can be relaxed to rapidly enough decaying fields at infinity, instead of compact support.}
Furthermore, the $n$-th derivative can be expressed as:
%------------------------------------------------
\begin{align}
    \frac{\dd^n}{\dd \lambda^n} S[\bar{\mathrm{Q}} + \lambda  \mathrm{q}] = \left[ \int \dd^\dimM x\  \mathrm{q}^I(x) \frac{\delta}{\delta \bar{\mathrm{Q}}^{I} (x) }\right]^n S[\bar{\mathrm{Q}} + \lambda  \mathrm{q}]
    \label{Eq:Partial_Derivatives_Gen}
\end{align}
%-------------------------------------------------------
where we are using the simplified notation \eqref{Eq:Shorcut_Notation} for the operator in the right hand side. Actually, it is possible to express all the terms $S^{(n)}$ for $n>2$ in terms of derivatives of $S^{(2)}[\bar{\mathrm{Q}},  \mathrm{q}]$. This is clear from Eqs.~\eqref{Eq:Partial_Actions_Gen} and~\eqref{Eq:Partial_Derivatives_Gen}, since we can combine them to find out that
%-------------------------------------------------------
\begin{align}
    S^{(n)} [\bar{\mathrm{Q}},  \mathrm{q}] &= \frac{1}{n!} \frac{\dd^n}{\dd \lambda^n} S [\bar{\mathrm{Q}} + \lambda  \mathrm{q}] \bigg\rvert_{\lambda  = 0}\nonumber\\
    &= \frac{1}{n!} \left[ \int \dd^\dimM x\  \mathrm{q}^I(x) \frac{\delta}{\delta \bar{\mathrm{Q}}^{I} (x) }\right]^n S[\bar{\mathrm{Q}} + \lambda  \mathrm{q}] \bigg\rvert_{\lambda  = 0}\nonumber\\
   !&= \frac{1}{n!} \left[ \int \dd^\dimM x\  \mathrm{q}^I(x) \frac{\delta}{\delta \bar{\mathrm{Q}}^{I} (x) }\right]^{n-2} \left(\left[ \int \dd^\dimM x\  \mathrm{q}^I(x) \frac{\delta}{\delta \bar{\mathrm{Q}}^{I} (x) }\right]^2 S[\bar{\mathrm{Q}} + \lambda  \mathrm{q}] \bigg\rvert_{\lambda  = 0}\right)\nonumber\\
    &= \frac{2}{n!} \left[ \int \dd^\dimM x\  \mathrm{q}^I(x) \frac{\delta}{\delta \bar{\mathrm{Q}}^{I} (x) }\right]^{n-2} S^{(2)} [\bar{\mathrm{Q}},  \mathrm{q}]\, .
\label{Eq:Quadratic_Actions}
\end{align}
%-------------------------------------------------------
We have used that the background $\bar{\mathrm{Q}}$ is generic, namely no information is lost when going from $S[\mathrm{Q}]$ to $S[\bar{\mathrm{Q}}]$. Having said this, notice that, once we move all the dynamics from $\mathrm{Q}^I$ to $ \mathrm{q}^I$, we have that $S^{(2)}$ is all we need to build out the whole action for $ \mathrm{q}^I$: $S^{(0)}$ is independent of $ \mathrm{q}^{I}$,\footnote{Hence, it is irrelevant for the classical dynamics.} and $S^{(1)}$ vanishes for a background $\bar{\mathrm{Q}}^{I}$ that is solution of the theory, {\it i.e.}, a configuration such that
%-------------------------------------------------------
\begin{equation}
    \frac{\delta S[\mathrm{Q}]}{\delta \mathrm{Q}^I} \bigg\rvert_{\mathrm{Q}^I=\bar{\mathrm{Q}}^I} = 0\,.
\end{equation}
%-------------------------------------------------------

%-------------------------------------------------------
\subsection{A functional identity}
\label{App:General_Fields_funidentity}
%-------------------------------------------------------

Now we proceed to show that each order in the expansion $S^{(n)}$ corresponds to the coupling of the $ \mathrm{q}^I$ field to the variation of $S^{(n-1)}$ with respect to the background of that field $\bar{\mathrm{Q}}^{I}$. Indeed, the following identity holds
%-------------------------------------------------------
\begin{align}
    \frac{\delta S^{(n)}   [\bar{\mathrm{Q}}, \mathrm{q}]}{\delta  \mathrm{q}^{I}} = \frac{\delta S^{(n-1)} [\bar{\mathrm{Q}}, \mathrm{q}]}{\delta \bar{\mathrm{Q}}^{I}}.
    \label{Eq:Key_Bootstrap_Gen}
\end{align}
%--------------------------------------------------------

%------------------------------------------------
\paragraph*{\textbf{Proof 1.}}
%------------------------------------------------
To show this identity we will follow~\cite{Butcher2009}. Let us first consider an arbitrary variation with respect to $\mathrm{q}^I$ or $\bar{\mathrm{Q}}^I$ and let us call it $X^I$. We have the following identity
%--------------------------------------------------------
\begin{align}
    \int \dd^\dimM x\ X^{I} (x) \frac{\delta S^{(n)}[\bar{\mathrm{Q}},\mathrm{q}]}{\delta \mathrm{q}^I (x)} = \int \dd^\dimM x\ X^{I} (x) \frac{1}{n!} \frac{\delta }{\delta \mathrm{q}^I (x) } \left(\frac{\dd^n}{\dd \lambda^n} S[\bar{\mathrm{Q}} + \lambda \mathrm{q}] \bigg\rvert_{\lambda  = 0}\right).
\end{align}
%--------------------------------------------------------
We can rewrite the last term as a derivative with respect to a new parameter $\sigma$.
%--------------------------------------------------------
\begin{align}
    \frac{1}{n!} \frac{\partial}{\partial \sigma } \frac{\partial^n}{\partial \lambda^n} S[\bar{\mathrm{Q}} + \lambda (\mathrm{q} + \sigma X) ] \bigg\rvert_{\lambda  = 0, \sigma = 0},
\end{align}
%--------------------------------------------------------
where clearly the derivatives are taken first and the substitution $\lambda = \sigma = 0$ is made at the end. Interchanging the order of derivatives
%--------------------------------------------------------
\begin{align}
    \frac{1}{n!} \frac{\partial^n}{\partial \lambda^n} \frac{\partial}{\partial \sigma } S[\bar{\mathrm{Q}} + \lambda (\mathrm{q} + \sigma X) ] \bigg\rvert_{\lambda  = 0, \sigma = 0}.
\end{align}
%--------------------------------------------------------
We can introduce a new variable $\tau = \lambda \sigma $, so that $\frac{1}{\lambda } \frac{\partial}{\partial \sigma} = \frac{\partial}{\partial \tau}$ at fixed $\lambda$. Hence, the expression above can be rewritten as
%--------------------------------------------------------
\begin{align}
    \frac{1}{n!} \frac{\partial^n}{\partial \lambda^n} \left(  \lambda \frac{\partial}{\partial \tau } S[\bar{\mathrm{Q}} + \lambda \mathrm{q} + \tau X ]\right) \bigg\rvert_{\lambda  = 0, \tau = 0} .
\end{align}
%--------------------------------------------------------
Evaluating the derivative with respect to $\lambda$ we find
%--------------------------------------------------------
\begin{align}
    & \frac{1}{n!} \left(\lambda \frac{\partial^n}{\partial \lambda^n}   \frac{\partial}{\partial \tau } S[\bar{\mathrm{Q}} + \lambda \mathrm{q} + \tau X]  + n \frac{\partial^{n-1}}{\partial \lambda^{n-1}} \frac{\partial}{\partial \tau} S [\bar{\mathrm{Q}} + \lambda \mathrm{q} + \tau X]\right) \bigg\rvert_{\lambda  = 0, \tau = 0} \nonumber \\
    &\qquad\qquad\qquad = \frac{1}{(n-1)!} \frac{\partial^{n-1}}{\partial \lambda^{n-1}} \frac{\partial}{\partial \tau} S [\bar{\mathrm{Q}} + \lambda \mathrm{q} + \tau X] \bigg\rvert_{\lambda  = 0, \tau = 0}.
\end{align}
%--------------------------------------------------------
Using the definition of $S_n[\bar{\mathrm{Q}},\mathrm{q}]$ introduced in Eq.~\eqref{Eq:Partial_Actions_Gen}, we find
%--------------------------------------------------------
\begin{align}
    \frac{\partial}{\partial \tau} S^{(n-1)} [\bar{\mathrm{Q}} + \tau X, \mathrm{q}] \bigg\rvert_{\tau = 0} = \int \dd^\dimM x\ X^I(x) \frac{\delta}{\delta \bar{\mathrm{Q}}^{I} (x)} S^{(n-1)} [\bar{\mathrm{Q}}, \mathrm{q}],
\end{align}
%--------------------------------------------------------
concluding our proof of Eq.~\eqref{Eq:Key_Bootstrap_Gen}. In all these steps we have ignored boundary terms as we mentioned above that we are assuming compact support variations.

%------------------------------------------------
\paragraph*{\textbf{Proof 2.}}
%------------------------------------------------
A perhaps more transparent derivation of this result can be given as follows. Let us begin with the definition of $S^{(n)}[\bar{\mathrm{Q}},\mathrm{q}]$
%--------------------------------------------------------
\begin{align}
    S^{(n)} [\bar{\mathrm{Q}},\mathrm{q}] = \frac{1}{n!} \left[ \int \dd^\dimM x\ \mathrm{q}^{I} (x) \frac{\delta}{\delta \bar{\mathrm{Q}}^I (x)} \right]^n S[\bar{\mathrm{Q}} + \lambda \mathrm{q}] \bigg\rvert_{\lambda  = 0} .
\end{align}
%--------------------------------------------------------
Now, let us compute the variation with respect to one of the fields $\mathrm{q}^J(y)$. For that purpose, we recall that the $n$ variations with respect to the background field can be rearranged using Eq.~\eqref{Eq:Shorcut_Notation}. Hence if we perform a variation with respect to $\mathrm{q}^J{x}$, it simply hits the product of the $n$-$\mathrm{q}^I(x)$ terms. This gives
%------------------------------------------------
\begin{align}
   & \frac{\delta}{\delta \mathrm{q}^{J}(y)} \int \dd^\dimM x_1 \ldots \int \dd^\dimM x_n \  \mathrm{q}^{I_1}(x_1) \ldots  \mathrm{q}^{I_n}(x_n) \frac{\delta}{\delta \bar{\mathrm{Q}}^{I_1}(x_1) } \ldots \frac{\delta}{\delta \bar{\mathrm{Q}}^{I_{n}}(x_n) } S [\bar{\mathrm{Q}} + \lambda \mathrm{q}]  \bigg\rvert_{\lambda  = 0}  \nonumber \\
   & \qquad = n \int \dd^\dimM z\ \delta^{\dimM } (y - z) \frac{\delta}{\delta \bar{\mathrm{Q}}^J (z)} \int \dd^\dimM x_1 \ldots \int \dd^\dimM x_{n-1} \  \mathrm{q}^{I_1}(x_1) \ldots  \mathrm{q}^{I_{n-1}}(x_{n-1}) \nonumber \\
   & \qquad \qquad \frac{\delta}{\delta \bar{\mathrm{Q}}^{I_1}(x_1) } \ldots \frac{\delta}{\delta \bar{\mathrm{Q}}^{I_{n-1}}(x_{n-1}) } S [\bar{\mathrm{Q}} + \lambda \mathrm{q}]  \bigg\rvert_{\lambda  = 0} .
\end{align}
%------------------------------------------------
The last part is the partial action $S^{(n-1)} [\bar{\mathrm{Q}},\mathrm{q}]$, up to a multiplicative factorial $(n-1)!$. This, together with the prefactor $n$ gives $n (n-1)! = n!$, cancelling the $n!$ above. Also, the $\delta$-distribution can be used to rearrange everything as simply a derivative with respect to $\bar{\mathrm{Q}}^J (y)$.
%------------------------------------------------
\begin{align}
    & \int \dd^\dimM z\ \delta^{\dimM } (y - z) \frac{\delta}{\delta \bar{\mathrm{Q}}^J (z)} \left[ \int \dd^\dimM x\ \mathrm{q}^{I} (x) \frac{\delta}{\delta \bar{\mathrm{Q}}^I (x)} \right]^{n-1} S[\bar{\mathrm{Q}} + \lambda \mathrm{q}] \bigg\rvert_{\lambda  = 0} \nonumber\\
    & \qquad = \int \dd^\dimM z \delta^{\dimM } (y - z) \frac{\delta}{\delta \bar{\mathrm{Q}}^J (z)} S^{(n-1)} [\bar{\mathrm{Q}},\mathrm{q}] \nonumber\\
    & \qquad = \frac{\delta}{\delta \bar{\mathrm{Q}}^J (y)} S^{(n-1)} [\bar{\mathrm{Q}},\mathrm{q}].
\end{align}
%------------------------------------------------
Hence, we find again expression~\eqref{Eq:Key_Bootstrap_Gen}.

%------------------------------------------------
\bibliographystyle{jhep}
\bibliography{bootstrap_biblio}
%------------------------------------------------

%------------------------------------------------
\end{document}